\documentclass[11pt]{article}
\usepackage[utf8]{inputenc}
\usepackage{graphicx}
\usepackage[colorlinks=true, urlcolor=citecolor, linkcolor=citecolor, citecolor=citecolor]{hyperref}

\usepackage{url}
\usepackage{dirtytalk} % for quotation
\usepackage{xspace} % removes extraneous spaces after user-defined commands
\usepackage{subfig}
\usepackage{float}
\usepackage[round]{natbib}
\usepackage{setspace}
\usepackage{float}

\usepackage[a4paper,left = 1in, right = 1in, bmargin = 1in]{geometry}
\usepackage{lineno}
\usepackage{authblk} % to add authors and affiliations

\usepackage[ruled,vlined]{algorithm2e}
\include{pythonlisting}
\usepackage{algpseudocode}

\usepackage{amsfonts, amsmath}
\usepackage{bm}
\DeclareMathOperator*{\argmaxA}{arg\,max} 
\newcommand{\ie}{i.e., }
\newcommand{\eg}{e.g., }
\newcommand{\TB}{TB\xspace}
\newcommand{\SB}{SB\xspace}
\newcommand{\supple}{supplementary material\xspace}
\newcommand{\manu}{main manuscript\xspace}

\providecommand{\keywords}[1]
{
  \small	
  \textbf{\textit{Keywords---}} #1
}

\title{Tailored Bayes: a risk modelling framework under unequal misclassification costs}

\vspace{-0.9cm}

\author[1,2]{Solon Karapanagiotis}
\author[3]{Umberto Benedetto}
\author[1,4]{Sach Mukherjee}
\author[1]{Paul D. W. Kirk}
\author[1]{Paul J. Newcombe}

\affil[1]{MRC Biostatistics Unit, University of Cambridge}
\affil[2]{The Alan Turing Institute}
\affil[3]{Bristol Heart Institute, University of Bristol}
\affil[4]{German Center for Neurodegenerative Diseases (DZNE), Germany}

\date{}

\begin{document}

% Running headers of paper:
%\markboth%
% First field is the short list of authors
%{S. Karapanagiotis and others}
% Second field is the short title of the paper
%{Tailored Bayes for risk prediction}

\maketitle

% Add a footnote for the corresponding author if one has been
% identified in the author list
%\footnote{Corresponding author: solon.karapanagiotis@mrc-bsu.cam.ac.uk}

\begin{abstract}{
Risk prediction models are a crucial tool in healthcare. Risk prediction models with a binary outcome (\ie binary classification models) are often constructed using methodology which assumes the costs of different classification errors are equal. In many healthcare applications this assumption is not valid, and the differences between misclassification costs can be quite large. For instance, in a diagnostic setting, the cost of misdiagnosing a person with a life-threatening disease as healthy may be larger than the cost of misdiagnosing a healthy person as a patient. In this work, we present Tailored Bayes (\TB), a novel Bayesian inference framework which \say{tailors} model fitting to optimise predictive performance with respect to unbalanced misclassification costs. %The method achieves this through the construction of datapoint-specific likelihood weights. %The method is versatile, as it allows for flexible construction of the weights. 
%We use simulation studies to showcase when \TB is expected to outperform standard Bayesian methods. 
We use simulation studies to showcase when \TB is expected to outperform standard Bayesian methods in the context of logistic regression. We then apply \TB to three real-world applications, a cardiac surgery, a breast cancer prognostication task and a breast cancer tumour classification task, and demonstrate the improvement in predictive performance over standard methods. 
}
\end{abstract}


%%%%%%%%%%%%%%%%%%%%%%%%%%%%%%%%%%%%%%%%%%%%%%%%%%%%%%%%%%%%%%%%%%%%%%%%%%%%%%%%%%
%%%%%%%%%%%%%%%%%%%%%%%%%%%%%%%%%%%%%%%%%%%%%%%%%%%%%%%%%%%%%%%%%%%%%%%%%%%%%%%%%%
%%%%%%%%%%%%%%%%%%%%%%%%%%%%%%%%%%%%%%%%%%%%%%%%%%%%%%%%%%%%%%%%%%%%%%%%%%%%%%%%%%
\section{Introduction}\label{introduction}
Risk prediction models are widely used in healthcare \citep{roques2003logistic, hippisley2008predicting, wishart2012predict}. In both diagnostic and prognostic settings, risk prediction models are regularly developed, validated, implemented, and updated with the aim of assisting clinicians and individuals in estimating probabilities of outcomes of interest which may ultimately guide their decision making \citep{down2014effect, nice2016cardiovascular, baumgartner2017}. The most common type of risk prediction model is based on binary outcomes, with class labels 0 (negative) and 1 (positive). Models for binary outcomes are often constructed to minimise the expected classification error; that is the proportion of incorrect classifications \citep{zhang2004statistical, steinwart2005consistency, bartlett2006convexity}. We refer to this paradigm as the standard classification paradigm. 

The disadvantage of this paradigm is that it implicitly assumes that all classification errors have equal costs, \ie the cost of misclassification of a positive label equals the cost of misclassification of a negative label. (Throughout the document we refer to the costs of incorrect classifications as misclassification costs). However, equal costs may not always be appropriate, and will depend on the scientific or medical context. For example, in cancer diagnosis, a false negative (that is, misdiagnosing a cancer patient as healthy) could have more severe consequences than a false positive (that is, misdiagnosing a healthy individual with cancer); the latter may lead to extra medical costs and unnecessary anxiety for the individual but not result in loss of life\footnotemark. For such applications, a prioritised control of asymmetric misclassification costs is needed.

\footnotetext{Note this example constitutes a simplification of the problem aimed to introduce the main idea of the paper, \ie in some applications the false positives and negatives have different costs. Hence, we are not considering the negative effects of toxicity of chemotherapy, overdiagnosis/unnecessary treatment for certain cancers, quality of life issues, etc.}

To meet this need, different methods have been developed. In the machine learning literature they are studied under the term cost-sensitive learning \citep{elkan2001foundations}. Existing research on cost-sensitive learning can be grouped into two main categories: direct and indirect approaches. Direct approaches aim to make particular classification algorithms cost-sensitive by incorporating different misclassification costs into the training process. This amounts to changing the objective/likelihood function that is optimised when training the model (\eg \cite{kukar1998cost, ling2004decision, masnadi2010risk}). A limitation is that these approaches are designed to be problem-specific, requiring considerable knowledge of the model in conjunction with its theoretical properties, and possibly new computational tools. Conversely, indirect approaches are more general because they achieve cost-sensitivity without any, or with minor modifications to existing modelling frameworks. In this work we focus on indirect approaches. %demonstrating their flexibility through applications to several clinical areas.

Indirect methods can be further subdivided into thresholding and sampling/weighting. Thresholding is the simplest approach of the two, as it changes the classification threshold of an existing risk prediction model. We can use the threshold to classify datapoints into positive or negative status if the model can produce probability estimates. This strategy is optimal if the true class probabilities were available. In other words, if the model is based on the logarithm of the ratio of true class probabilities, the threshold should be modified by a value equal to the logarithm of the ratio of misclassification costs \citep{duda2012pattern}. This is based on decision theoretic arguments as we show in Section \ref{methods} \citep{pauker1975therapeutic, duda2012pattern}. In practice, however, this strategy may lead to sub-optimal solutions. We demonstrate this using synthetic (Section \ref{simulations}) and real-life data (Section \ref{real_data}). 

Alternatively, sampling methods modify the distribution of the training data according to misclassification costs (see \cite{elkan2001foundations} for a theoretical justification). This can be achieved by generating new datapoints from the class with smaller numbers of datapoints \ie oversampling from the minority class,
%the minority class (oversampling) 
or by removing datapoints from the majority class (undersampling). The simplest form is random sampling (over- or under-). However, both come with drawbacks. Duplicating samples from the minority class may cause overfitting \citep{zadrozny2003cost}. Similarly, random elimination of samples from the majority class can result in loss of data which might be useful for the learning process. 
Weighting (\eg \cite{ting1998inducing, margineantu2003wrapper}) can also be conceptually viewed as a sampling method, where weights are assigned proportionally to misclassification costs. For example, datapoints of the minority class, which usually carries a higher misclassification cost, may be assigned higher weights. Datapoints with high weights can be viewed as sample duplication – thus oversampling. In general, random sampling/weighting determine the datapoints to be duplicated or eliminated based on outcome information (whether a datapoint belongs to the majority or the minority class). Notably, they do not take into account critical regions of the covariate space, such as regions that are closer to the target decision boundary. A decision boundary specifies distinct classification regions on the covariate space based on specified misclassification costs (see Section \ref{simulations} for details). This is the goal of the framework presented here.

In this paper, we build upon the seminal work of \cite{hand2003local}, and present an umbrella framework that allows us to incorporate misclassification costs into commonly used models for binary outcomes. The framework allows us to tailor model development with the aim of improving performance in the presence of unequal misclassification costs.
%In this paper, we build upon the seminal work of \cite{hand2003local} and present an umbrella framework that allows us to tailor most commonly used models for binary classification. The framework targets model development which performs well under unequal misclassification costs.
Although the concepts we discuss are general, and allow for relatively simple tailoring of a wide range of models (essentially whenever the objective function can be expressed as a sum over samples), we focus on a Bayesian regression paradigm.
%Although the concepts are general, we concentrate on a Bayesian paradigm. 
Hence, we present Tailored Bayes (\TB), a framework for tailored Bayesian inference when different classification errors incur different penalties. 
%In  this  paper  we  present  Tailored  Bayes  (\TB),  a  framework for  tailored  Bayesian  inference  when different classification errors incur different penalties. Our framework targets model development which performs well under unequal misclassification costs. Although the concepts are general, we concentrate on Bayesian approaches.
We use a decision theoretic approach to quantify the benefits and costs of correct and incorrect classifications (Section \ref{methods}). The method is based on the principle that the relative harms of false positives and false negatives can be expressed in terms of a target threshold. We then build a 2-stage model (Section \ref{model}); first introduced by \cite{hand2003local}. In the first stage, the most informative datapoints are identified. A datapoint is treated as informative if it is close to the target threshold of interest. Each datapoint is assigned a weight proportional to its distance from the target threshold. Intuitively, one would expect improvements in performance to be possible by putting decreasing weights on the class labels of the successively more distant datapoints. In the second stage, these weights are used to downweight each datapoint's likelihood contribution during model training. A key feature is that this changes the estimation output in a way that goes beyond thresholding and we demonstrate this effect in simple examples (Section \ref{simulations}).

We conduct simulation studies to illustrate the improvement in predictive performance of our proposed \TB modelling framework over the standard Bayesian paradigm (Section \ref{simulations}). We then apply the methodology to three real-data applications (Section \ref{real_data}). Our two main case studies are a breast cancer and a cardiac surgery prognostication task where we have information on how clinicians prioritise different classification errors. We show that incorporating this information into the model through our \TB approach leads to better treatment decisions. We finish with a discussion of our approach, findings and provide some general remarks in Section \ref{discussion}. 
%%%%%%%%%%%%%%%%%%%%%%%%%%%%%%%%%%%%%%%%%%%%%%%%%%%%%%%%%%%%%%%%%%%%%%%%%%%%%%%%%%
%%%%%%%%%%%%%%%%%%%%%%%%%%%%%%%%%%%%%%%%%%%%%%%%%%%%%%%%%%%%%%%%%%%%%%%%%%%%%%%%%%
%%%%%%%%%%%%%%%%%%%%%%%%%%%%%%%%%%%%%%%%%%%%%%%%%%%%%%%%%%%%%%%%%%%%%%%%%%%%%%%%%%
\section{Methods}\label{methods}
We use a decision theoretic approach to summarise the costs of misclassifications of a binary outcome into a single number, which we refer to as the target threshold (Section \ref{concepts}). We later (Section \ref{net_benefit_risk_prediction})  define the expected utility of risk prediction and use the target threshold and the never treat policy to simplify the expected utility and derive the Net Benefit of a risk prediction model. We use the Net Benefit as our model evaluation metric throughout the paper. In Section \ref{model} we incorporate the target threshold in the model formulation which results in the tailored likelihood function (Section \ref{tailored_likelihood_function}) and the tailored posterior (Section \ref{bayesian_tailoring}).

%%%%%%%%%%%%%%%%%%%%%%%%%%%%%%%%%%%%%%%%%%%%%%%%%%%%%%%%%%%%%%%%%%%%%%%%%%%%%%%%%%
\subsection{The target threshold}\label{concepts}
%In this section we define the target threshold, a scalar function of the four basic utilities of prediction. 

%we use a decision theoretic approach to summarise the costs of misclassifications of a binary outcome into a single number, which we refer to as the target threshold. %We later use the target threshold to define the net benefit of a risk prediction model, which is our target performance measure. 
Let $Y \in \{0, 1\}$ represent a binary outcome of interest. The observed $Y$ is a realisation of a binary random variable following a Bernoulli distribution with $\pi = P[Y = 1]$. This is the marginal probability of the outcome being present, and consequently, the probability the outcome being absent is $(1 - \pi)$.

We introduce utility functions to take into account the benefits or harms of different classifications. A utility function assigns a value to each of the four possible classification-outcome combinations stating exactly how beneficial/costly each action (treat or no treat) is. We assume that people who are classified as positive receive treatment and people who are classified as negative do not receive treatment. We use \say{treatment} in the generic sense of healthcare intervention which could be a drug, surgery or further testing. Each possible combination of classification (negative and positive) and outcome status (0, 1) is associated with an utility where a positive value indicates a benefit and a negative value indicates a cost or harm. The four utilities associated with binary classification problems are: 
(a) $U_{TP}$, the utility of a true positive classification, that is administering treatment to a patient who has the outcome (\ie treat when necessary), (b) $U_{FP}$, the utility of a false positive classification, that is the utility of administering treatment to a patient who does not have the outcome (\ie administering unnecessary treatment), 
(c) $U_{FN}$, the utility of a false negative classification, that is the utility of withholding treatment from a patient that has the outcome (\ie withholding beneficial treatment), and (d) $U_{TN}$, the utility of a true negative classification, that is the utility of withholding treatment from a patient who does not have the outcome (\ie withholding unnecessary treatment).

%\begin{itemize}
%    \item $U_{TP}$, is the utility of a true positive classification, that is administering treatment to a patient who has the outcome (\ie treat when necessary),
%    \item $U_{FP}$, is the utility of a false positive classification, that is the utility of administering treatment to a patient who does not have the outcome (\ie administering unnecessary treatment), 
 %   \item $U_{FN}$, is the utility of a false negative classification, that is the utility of withholding treatment from a patient that has the outcome (\ie withholding beneficial treatment), and 
%    \item $U_{TN}$, is the utility of a true negative classification, that is the utility of withholding treatment from a patient who does not have the outcome (\ie withholding unnecessary treatment). 
%\end{itemize}

The expected utilities of the two fixed courses of action (or policies) of always treat and never treat are given by
%The expected utilities of both courses of action (or policies) (treat and not treat) are given by
\begin{subequations}
    \begin{equation} \label{EU_treat}
    %\begin{split}
        EU_{treat} =  \pi U_{TP} + (1 - \pi) U_{FP},%\\
    \end{equation}
    \begin{equation} \label{EU_notreat}
        EU_{no \text{ } treat} =  \pi U_{FN} + (1 - \pi) U_{TN}
    %\end{split}
\end{equation}
\end{subequations}

where $EU_{treat}$ and $EU_{no \text{ } treat}$ are the expected utility of treating and not treating, respectively.
In principle, one should choose the course of action with the highest expected utility. When the expected utilities are equal, the decision maker is indifferent on the course of action \citep{pauker1975therapeutic}. 
Based on classical decision theory, we employ the threshold concept and denote with $t$ the threshold at which the decision maker is indifferent on the course of action \citep{pauker1980threshold}. This is the principle of clinical equipoise which exists when all of the available evidence about a course of action does not show that it is more beneficial than an alternative and, equally, does not show that it is less beneficial than the alternative \citep{Turner2013}. Clinical equipoise is regarded as an \say{ethically necessary condition in all cases of clinical research} \citep{freedman1987equipoise}. Based on the threshold concept, an individual should be treated (\ie classified as positive) if $\pi \geq t$ and should not be treated (\ie classified as negative) otherwise. Having defined $t$ as the value of $\pi$ of clinical equipoise where the expected benefit of treatment is equal to the expected benefit of avoiding treatment implies $EU_{treat} =  EU_{no \text{ } treat}$ or equivalently,  $t U_{TP} + (1 - t) U_{FP} = t U_{FN} + (1 - t) U_{TN}$. Solving for $t$, 
\begin{equation}
    \begin{split}
    t = \frac{U_{TN} - U_{FP}}{U_{TN} - U_{FP} + U_{TP} - U_{FN}}\\
        = \frac{H}{H + B} = \frac{1}{1 + \frac{B}{H}},
        \label{target_threshold}
    \end{split}
\end{equation}        

where $B = U_{TP} - U_{FN}$ is the difference between the utility of administering treatment to individuals who have the outcome and the utility of withholding treatment in those who have the outcome. In other words, $B$ is the benefit for positive prediction, and consequent treatment, among those with the outcome. Similarly, $B$ can be interpreted as the consequence of failing to treat when it would have been of benefit, that is, the harm from a false negative result (compared to a true positive result). Comparably, $H$ is the difference between the utility of avoiding treatment in patients who do not have the outcome and the utility of administering treatment to those who do not have the outcome (\ie $U_{TN} - U_{FP}$). In other words, $H$ is the consequence of being treated unnecessarily, this is the harm associated with a false positive result (compared to a true negative result). 

We henceforth refer to $t$ as the target threshold. Alternative names in the literature are risk threshold \citep{baker2009using} and threshold probability \citep{tsalatsanis2010regret}. It is a scalar function of $U_{TP},U_{FN}, U_{TN}$ and $U_{FP}$ that determines the cut-off point for calling a result positive that maximizes expected utility. Equation (\ref{target_threshold}) therefore tells us that the target threshold at which the decision maker will opt for treatment is informative of how they weigh the relative harms of false positive and false negative results. The main advantage of this decision theoretic approach is there is no need to explicitly specify the relevant utilities, but only the desired target threshold.

\begin{quotation}
{\bf Example:} %Assume that we are willing to treat no more than 10 patients in order to have one true positive, \ie one correctly treated patient. 
Assume that for every correctly treated patient (true positive) we are willing to incorrectly treat 9 healthy individuals (false positives)\footnotemark.
Then we consider the benefit of correctly treating a patient to be nine times larger than the harm of an unnecessary treatment: the harm-to-benefit ratio is 1:9. This ratio has a direct relationship to $t$: the odds of $t$ equal the harm-to-benefit ratio. That is, $H/B = t / (1 - t)$ which is implied by (\ref{target_threshold}). For example, $t$ of 10\% implies a harm-to-benefit ratio of 1:9 (odds(10\%) = 10/90). 
\end{quotation}

\footnotetext{The statement is equivalent to the following:
Assume that not treating an individual with the outcome (false negative) is 9 times worse than treating unnecessarily a healthy individual (false positive). Both statements result in the same harm-to-benefit ratio.}

%%%%%%%%%%%%%%%%%%%%%%%%%%%%%%%%%%%%%%%%%%%%%%%%%%%%%%%%%%%%%%%%%%%%%%%%%%%%%%%%%%
\subsection{Net Benefit for risk prediction}\label{net_benefit_risk_prediction}
In practice, we do not know the probability of the outcome of any given individual. Instead, we need to estimate it, according to a set of covariates. Let $\mathbf{X} \in \mathbb{R}^d$ be a vector of $d$ covariates and define $\pi(\mathbf{x})$ as the conditional class 1 probability given the observed values of the covariates, $\mathbf{x}: \pi(\mathbf{x}) = P[Y = 1 \mid \mathbf{X} = \mathbf{x}]$.
We are concerned with the problem of classifying future values of $Y$ from the information that the covariates $\mathbf{X}$ contain. 
%More precisely, we classify as positive if $\pi(\mathbf{x}) \ge t$, and as negative otherwise.
Assume we have a prediction model and an estimate of $\pi(\mathbf{x})$, denoted $\hat{\pi}(\mathbf{x})$. We classify an individual as positive if $\hat{\pi}(\mathbf{x}) \ge t$, where $t$ is the target threshold (defined in (\ref{target_threshold})) and as negative otherwise. The expected utility of assigning treatment or not (\ie classifying positive or negative) at $t$  based on the model's predictions $\hat{\pi}(\mathbf{x})$ can be written as
\begin{equation}
    \begin{split}
        EU_{Pred(t)} = P(\hat{\pi}(\mathbf{x}) \geq t, y = 1) U_{TP} + P(\hat{\pi}(\mathbf{x}) < t, y = 1) U_{FN} + \\
        P(\hat{\pi}(\mathbf{x}) < t, y = 0) U_{TN} +  
        P(\hat{\pi}(\mathbf{x}) \geq t, y = 0) U_{FP} \\
       = \pi  TPR_t  U_{TP} + \pi  (1 - TPR_t) U_{FN} + (1 - \pi)  FPR_t  U_{FP} + (1 - \pi)  (1 – FPR_t)  U_{TN}\\
       = \{\pi TPR_t B  - (1 - \pi) FPR_t H \} + \{\pi U_{FN} + (1 - \pi) U_{TN} \},
    \end{split}
    \label{full_EU}
\end{equation} 

where $TPR_t$ is the true positive rate, \ie $P(\hat{\pi}(\mathbf{x}) \geq t|y = 1)$ and $FPR_t$ is the false positive rate, \ie $P(\hat{\pi}(\mathbf{x}) \geq t|y = 0)$. The drawback of this formulation is the need to specify the four utilities. Equation (\ref{full_EU}) can be simplified by considering the expected utility of risk prediction in excess of the expected utility of no treatment. The expected utility of no treatment is 
given in (\ref{EU_notreat}), and so, subtracting this from both sides of (\ref{full_EU}), the expected utility of risk prediction in excess of the expected utility of no treatment is
%\begin{equation} % first version 
%    \begin{split}
%        EU_{Pred(t)} - EU_{no \text{ } treat} = \pi TPR_t B - (1 - \pi) FPR_t H\\
%        = B \big\{ \pi  TPR_t - (1 - \pi) FPR_t \frac{t}{1 - t}\big\}.
%    \end{split}
%    \label{net_benefit_long}
%\end{equation}
\begin{equation}
\begin{array}{l@{}l}
EU_{Pred(t)} - EU_{no \text{ } treat} &{}= \pi TPR_t B - (1 - \pi) FPR_t H\\
    &{}= B \big\{ \pi  TPR_t - (1 - \pi) FPR_t \frac{t}{1 - t}\big\}.
\end{array}
\label{net_benefit_long}
\end{equation}

%This is a Hippocratic utility function because it is motivated by the Hippocratic oath. It incorporates both the principles of \textit{beneficence} (do the best in one’s ability) and
%\textit{non-maleficence} (do no harm). Both principles are considered central notions in bioethics \citep{childress2001principles}. To be consistent with the Hippocrates’s oath, the modeller chooses the model that has the greatest chance of giving an outcome no worse than the outcome of no treatment.
This is a Hippocratic utility function because it is motivated by the Hippocratic oath; do the best in one’s ability (beneficence) and do no harm (non-maleficence) \citep{childress2001principles}. To be consistent with the Hippocratic oath, the modeller chooses the model that has the greatest chance of giving an outcome no worse than the outcome of no treatment. With $B = 1$, (\ref{net_benefit_long}) is defined as the Net Benefit of risk prediction versus treat none \citep{vickers2006decision, baker2009using}. Setting $B = 1$ as the reference level means that Net Benefit is measured in units of true positive predictions. To see this we re-write (\ref{net_benefit_long}) as 
\begin{equation}
    NB_{Pred(t)} = \frac{TP_t}{n} - \frac{FP_t}{n} \frac{t}{1 - t},
 \label{NB_model}
\end{equation}

where $TP_t$ is number of patients with true positive results, $FP_t$ is number of patients with false positive results, and $n$ is the sample size. To simplify notation we write $NB$ instead of $NB_{Pred(t)}$. $NB$ gives the proportion of net true positives in the dataset, accounting for the different misclassification costs. In other words, 
the observed number of true positives is corrected for the observed proportion of false positives weighted by the odds of the target threshold, and the result is divided by the sample size. This net proportion is equivalent to the proportion of true positives in the absence of false positives. For instance, a $NB$ of 0.05 for a given target threshold, can be interpreted as meaning that use of the model, as opposed to simply assuming that all patients are negative, leads to the equivalent of an additional 5 net true positives per 100 patients.

For the remainder of the manuscript $NB$ will be our main performance measure for model evaluation. %In the above, we have shown it is defined as a function of the target threshold $t$, which captures the relative utilities of treatment decisions. In the next section we incorporate these utilities, through $t$, into the model formulation. 
We have written $NB$ as a function of the target threshold $t$, which allows information about the relative utilities of treatments to be included in our model formulation, which we now show.

%%%%%%%%%%%%%%%%%%%%%%%%%%%%%%%%%%%%%%%%%%%%%%%%%%%%%%%%%%%%%%%%%%%%%%%%%%%%%%%%%%
\subsection{Model formulation}\label{model}
Denote data $D = \{ (y_i, \mathbf{x}_i)  :i=  1, \dots, n \}$ where $y_i$ is the outcome indicating the class to which the $i^{th}$ datapoint belongs and $\mathbf{x}_i$ is the vector of covariates of size $d$. The objective is to estimate the posterior probability of belonging to one of the classes given a set of new datapoints. We use $D$ to fit a model $p(y_i \mid \mathbf{x}_i)$ and use it to obtain $\pi(\mathbf{x}_*)$ for a future datapoint $y_*$ with covariates $\mathbf{x}_*$. We simplify the structure using $p(y_i \mid f(\mathbf{x}_i))$, where $f: \mathcal{X} \to \mathbb{R}$ is a function that maps the vector of the covariates to the real line \ie the linear predictor used in generalised linear models. To develop the complete model, we need to specify $p(y_i \mid f(\mathbf{x}_i))$ and $f$.

In the machine learning literature, most of the binary classification procedures use a loss-function-based approach. In the same spirit, we model $p(y_i \mid f(\mathbf{x}_i))$ according to a loss function
$\ell(y_i, f(\mathbf{x}_i))$ which measures the loss for reporting $f$ when the truth is $y$. Mathematically, minimizing this loss function can be equivalent to maximizing $-\ell(y, f)$, where $\exp \{ -\ell(y, f)\}$ is proportional to the likelihood function. This duality between `likelihood' and `loss', that is viewing the loss as the negative of the log-likelihood is referred to in the Bayesian literature as a logarithmic score (or loss) function \citep{bernardo2009bayesian, bissiri2016general}. A few popular choices of loss functions for binary classification are the exponential loss used in boosting classifiers \citep{friedman2000additive}, the hinge loss of support vector machines \citep{zhang2004statistical}, or logistic loss of logistic regression \citep{friedman2000additive, zhang2004statistical}. In this work, we focus on the following loss, 
\begin{equation}
   \ell_{w_i}(y_i, f(\mathbf{x}_i)) = -\pi(f(\mathbf{x}_i))^{w_{i}y_i} (1- \pi(f(\mathbf{x}_i)))^{w_i(1 - y_i)}, \text{ for } i = 1, \ldots, n
   \label{hand_loss}
\end{equation}

where we define $\pi_{w_{i}}(f(\mathbf{x}_i)) := \pi(f(\mathbf{x}_i))^{w_i} =(\exp \{ {\mathbf{x}^T_i \bm{\beta}} \}/ 1 + \exp\{ {\mathbf{x}^T_i \bm{\beta}}\})^{w_i}$ and $w_i \in[0,1]$ are datapoint-specific weights. This is a generalised version of the logistic loss, first introduced by \cite{hand2003local}. 
We recover the standard logistic loss by setting $w_i = 1$ for all $i = 1, \dots ,n$. Note that we specify $f$ as a linear function, \ie $f(\mathbf{x}_i)= \mathbf{x}^T_i \bm{\beta}$, where $\bm{\beta}$ is a $d + 1$ dimensional vector of regression coefficients. Hence, our objective is to learn $\bm{\beta}$. We make this explicit by replacing $\pi_{w_{i}}(f(\mathbf{x}_i))$ with $\pi_{w_{i}}(\mathbf{x}_i;\bm{\beta})$ for the rest of this work.

The datapoint-specific weights, $w_i$, allow us to tailor the standard logistic model. We wish to weigh observations based on their vicinity to the target threshold, $t$, upweighting observations close to $t$ (the most informative) and downweighting those that are further away. To accomplish this we set the weights as 
\begin{equation}
    w_i = \exp \big \{-\lambda h(\pi_u(\mathbf{x}_i), t) \big \} = \exp \big \{-\lambda (\pi_u(\mathbf{x}_i) - t)^2 \big \},
    \label{weights}
\end{equation}

where $h$ is the squared distance (see \supple for other options) and $\pi_u(\mathbf{x}_i)$ is the unweighted version of $\pi_{w_{i}}(\mathbf{x}_i;\bm{\beta})$. Of course, in practice we do not know $\pi_u(\mathbf{x}_i)$ so we cannot measure the distance between $t$ and each datapoint's predicted probability, $\pi_u(\mathbf{x}_i)$, in order to derive these weights. To overcome this, we propose a two-stage procedure. First, the distance is measured according to an estimate of $\pi_u(\mathbf{x}_i)$, $\hat{\pi}_u(\mathbf{x}_i)$, which can be compared with $t$ to yield the weights. This estimate could be based on any classification method: we use standard unweighted Bayesian logistic regression in the analysis below. If a well-established model of $\pi_u(\mathbf{x}_i)$ already exists in the literature that could be used (as in our cardiac surgery case study, see Section \ref{Real_data_application_2}) this task would not be necessary. After deriving the weights, they are then used to estimate $\pi_{w_{i}}(\mathbf{x}_i;\bm{\beta})$. %Since in the analyses below we estimate both $\pi_u(\mathbf{x}_i)$ and $\pi_{w_{i}}(\mathbf{x}_i;\bm{\beta})$ from the same data, we use a data splitting approach to avoid overfitting (see Section \ref{data_splitting}). 
Finally, under the formulation in (\ref{weights}) the weights decrease with increasing distance from the target threshold $t$. The tuning parameter $\lambda \geq 0$ controls the rate of that decrease. For $\lambda = 0$ we recover the standard logistic regression model. We use cross-validation to choose $\lambda$, see later for details.

%%%%%%%%%%%%%%%%%%%%%%%%%%%%%%%%%%%%%%%%%%%%%%%%%%%%%%%%%%%%%%%%%%%%%%%%%%%%%%%%%%
\subsection{Tailored likelihood function}\label{tailored_likelihood_function}
To gain a better insight into the model we define the tailored likelihood function as 
\begin{equation}
    L(D \mid \bm{\beta}) = -\prod_{i=1}^{n} \ell_{w_i}(y_i, \mathbf{x}^T_i \bm{\beta}) = \prod_{i=1}^{n}  \Bigg( \frac{\exp \{ {\mathbf{x}^T_i \bm{\beta}}\}}{1 + \exp \{ {\mathbf{x}^T_i \bm{\beta}\}}}  \Bigg)^{y_iw_i} \Bigg( 1 - \frac{\exp \{ {\mathbf{x}^T_i \bm{\beta}}\}}{1 + \exp \{ {\mathbf{x}^T_i \bm{\beta}}\}} \Bigg)^{w_i (1-y_i)}
    \label{tailored_likelihood}
\end{equation}

Strictly speaking, this quantity is not the standard logistic likelihood function. Nevertheless, it is instinctive to see its correspondence with the standard likelihood function. Thus, we rewrite (\ref{tailored_likelihood}) (after taking the log in both sides) as

%\begin{equation}
%    \begin{split}
%    \log(L(D \mid \bm{\beta})) = -\sum_{i=1}^{n} \log(\ell_{w_i}(y_i, \mathbf{x}^T_i \bm{\beta}))\\ = 
%    \sum_{i=1}^{n}  {y_iw_i} \log\Bigg( \frac{\exp \{ {\mathbf{x}^T_i \bm{\beta}}\}}{1 + \exp \{ {\mathbf{x}^T_i \bm{\beta}\}}} \Bigg)  + {w_i (1-y_i)\log \Bigg( 1 - \frac{\exp \{ {\mathbf{x}^T_i \bm{\beta}}\}}{1 + \exp \{ {\mathbf{x}^T_i \bm{\beta}}\}} \Bigg)}\\ 
%    = \sum_{i=1}^{n} w_i \Bigg[ y_i \log\Bigg( \frac{\exp \{ {\mathbf{x}^T_i \bm{\beta}}\}}{1 + \exp \{ {\mathbf{x}^T_i \bm{\beta}\}}} \Bigg)  + {(1-y_i) \log \Bigg( 1 - \frac{\exp \{ {\mathbf{x}^T_i \bm{\beta}}\}}{1 + \exp \{ {\mathbf{x}^T_i \bm{\beta}}\}} \Bigg)}\Bigg] \\
%    = \sum_{i=1}^{n} w_i l_i(D \mid \bm{\beta}) 
%    \end{split}
%    \label{tailored_likelihood2}
%\end{equation}

\begin{equation}
\begin{array}{l@{}l}
\log(L(D \mid \bm{\beta})) &{} = -\sum_{i=1}^{n} \log(\ell_{w_i}(y_i, \mathbf{x}^T_i \bm{\beta})) \\
    &{}= \sum_{i=1}^{n}  {y_iw_i} \log\Bigg( \frac{\exp \{ {\mathbf{x}^T_i \bm{\beta}}\}}{1 + \exp \{ {\mathbf{x}^T_i \bm{\beta}\}}} \Bigg)  + {w_i (1-y_i)\log \Bigg( 1 - \frac{\exp \{ {\mathbf{x}^T_i \bm{\beta}}\}}{1 + \exp \{ {\mathbf{x}^T_i \bm{\beta}}\}} \Bigg)} \\
    &{}= \sum_{i=1}^{n} w_i \Bigg[ y_i \log\Bigg( \frac{\exp \{ {\mathbf{x}^T_i \bm{\beta}}\}}{1 + \exp \{ {\mathbf{x}^T_i \bm{\beta}\}}} \Bigg)  + {(1-y_i) \log \Bigg( 1 - \frac{\exp \{ {\mathbf{x}^T_i \bm{\beta}}\}}{1 + \exp \{ {\mathbf{x}^T_i \bm{\beta}}\}} \Bigg)}\Bigg] \\
    &{}= \sum_{i=1}^{n} w_i l_i(D \mid \bm{\beta})
\end{array}
\label{tailored_likelihood2}
\end{equation}

%\begin{multline}
%    \log(L(D \mid \bm{\beta})) = -\sum_{i=1}^{n} \log(\ell_{w_i}(y_i, \mathbf{x}^T_i \bm{\beta}))\\ 
%    = \sum_{i=1}^{n}  {y_iw_i} \log\Bigg( \frac{\exp \{ {\mathbf{x}^T_i \bm{\beta}}\}}{1 + \exp \{ {\mathbf{x}^T_i \bm{\beta}\}}} \Bigg)  + {w_i (1-y_i)\log \Bigg( 1 - \frac{\exp \{ {\mathbf{x}^T_i \bm{\beta}}\}}{1 + \exp \{ {\mathbf{x}^T_i \bm{\beta}}\}} \Bigg)}\\ 
%    = \sum_{i=1}^{n} w_i \Bigg[ y_i \log\Bigg( \frac{\exp \{ {\mathbf{x}^T_i \bm{\beta}}\}}{1 + \exp \{ {\mathbf{x}^T_i \bm{\beta}\}}} \Bigg)  + {(1-y_i) \log \Bigg( 1 - \frac{\exp \{ {\mathbf{x}^T_i \bm{\beta}}\}}{1 + \exp \{ {\mathbf{x}^T_i \bm{\beta}}\}} \Bigg)}\Bigg] \\ 
%    = \sum_{i=1}^{n} w_i l_i(D \mid \bm{\beta}) 
%    \label{tailored_likelihood2}
%\end{multline}

where $l_i(D \mid \bm{\beta})$ is the standard logistic log-likelihood function. We can further replace (\ref{weights}) into (\ref{tailored_likelihood2})
\begin{equation*}
\log(L(D \mid \bm{\beta})) = \sum_{i=1}^{n} \exp \big \{-\lambda (\pi_u(\mathbf{x}_i) - t)^2 \big \}  l_i(D \mid \bm{\beta})
\end{equation*}

to see that each datapoint contributes exponentially proportional to its distance from the target threshold $t$, which summarises the four utilities associated with binary classification problems (see \ref{target_threshold}). One option to proceed is by optimising  the tailored likelihood function with respect to the coefficients in an empirical risk minimisation approach \citep{vapnik1998statistical}. An attractive feature of (\ref{tailored_likelihood2}) is that this optimisation is computationally efficient since we can rely on existing algorithmic tools, \eg (stochastic) gradient optimisation. However, here we learn the coefficients in a Bayesian formalism. 
%%%%%%%%%%%%%%%%%%%%%%%%%%%%%%%%%%%%%%%%%%%%%%%%%%%%%%%%%%%%%%%%%%%%%%%%%%%%%%%%%%
\subsection{Bayesian tailoring}\label{bayesian_tailoring}
%We present a Bayesian version of the model described above. 
Following Bayes Theorem, the \TB posterior is 
\begin{equation}
p(\bm{\beta} \mid D) = \frac{L(D \mid \bm{\beta}) p(\bm{\beta})}{p(D)},
    \label{tailored_posterior_main}
\end{equation}
where $L(D \mid \bm{\beta})$ is the tailored likelihood function given in (\ref{tailored_likelihood}), $p(\bm{\beta})$ is the prior on the coefficients, and $p(D) = \int L(D|\tilde{\bm{\beta}}) p(\tilde{\bm{\beta}}) d\tilde{\bm{\beta}}$, is the normalising constant. In this work we assume a normal prior distribution for each element of $\bm{\beta}$, \ie $p(\beta_j) =  \mathcal{N}(\mu_j, \sigma_j^2),$ where $\mu_j$ and $\sigma_j$ are the mean and standard deviation respectively for the $j^{th}$ element of $\bm{\beta}$ ($j = 1, \dots, d + 1$). For all analysis below we use vague priors with $\mu_j= 0$ and $\sigma_j = 100$, for all $j$. 

Conveniently, we can interpret the choice of prior as a regularizer on a per-datapoint influence/importance (see \supple Section S1). Crucially, this allows us to view the \TB posterior as combining a standard likelihood function with a data-dependent prior (\supple Section S1). Hence, even though the tailored likelihood function does not have a probabilistic interpretation the \TB posterior is a proper posterior. 

In the \supple we provide details on the model inference and predictions steps (Section S2), the cross-validation scheme for choosing $\lambda$ (Section S3), the data-spitting strategy (Section S4), and the Markov chain Monte Carlo (MCMC) algorithm we are implementing (Section S5).

%%%%%%%%%%%%%%%%%%%%%%%%%%%%%%%%%%%%%%%%%%%%%%%%%%%%%%%%%%%%%%%%%%%%%%%%%%%%%%%%%%
%%%%%%%%%%%%%%%%%%%%%%%%%%%%%%%%%%%%%%%%%%%%%%%%%%%%%%%%%%%%%%%%%%%%%%%%%%%%%%%%%%
%%%%%%%%%%%%%%%%%%%%%%%%%%%%%%%%%%%%%%%%%%%%%%%%%%%%%%%%%%%%%%%%%%%%%%%%%%%%%%%%%%
\section{Simulations}\label{simulations}

The simulations are designed to provide insight into when \TB  can be advantageous compared to the standard Bayesian paradigm. Two scenarios where \TB is expected to outperform standard Bayes (\SB) are the absence of parallelism of the optimal decision boundaries and data contamination. A decision boundary determines distinct classification regions in the covariate space. It provides a rule to classify datapoints based on whether the datapoint's covariate vector falls inside or outside the classification region. If a datapoint falls inside the classification region it will be labelled as belonging to class 1 (\eg positive), if it falls outside it will be labelled as belonging to class 0 (\eg negative).
According to Bayesian decision theory the optimal decision boundaries determine the classification regions where the expected reward is maximised given pre-specified misclassification costs \citep{duda2012pattern}. More specifically, we classify as positive if $\frac{\pi(\mathbf{x})}{1- \pi(\mathbf{x})} > \frac{t}{1-t}$, where $\pi(\mathbf{x})$ denotes the true class 1 probability, as in Section \ref{concepts}. Simulations 1 and 2 present two settings where the optimal decision boundaries are not parallel with their orientation changing as a function of the target threshold. Simulation 3 is an example of data contamination.     

%%%%%%%%%%%%%%%%%%%%%%%%%%%%%%%%%%%%%%%%%%%%%%%%%%%%%%%%%%%%%%%%%%%%%%%%%%%%%%%%%%
\subsection{Simulation 1: Linear Decision Boundaries}\label{sim_1}
\begin{figure}[H]
	\begin{centering}
		\subfloat[a][]{\includegraphics[width=0.50\linewidth]{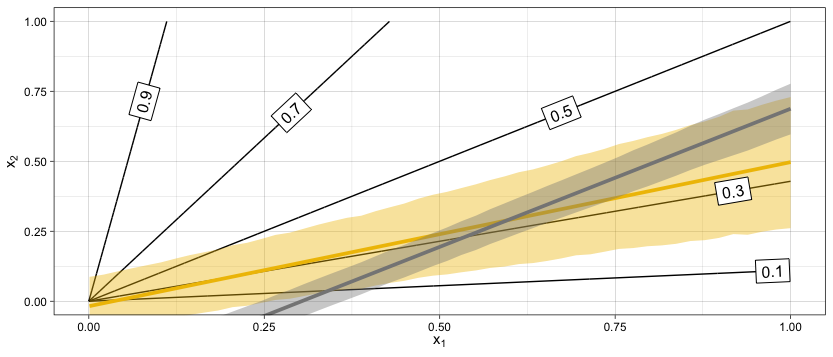}\label{fig:a_fig1_sim1}} 
		\subfloat[b][]{\includegraphics[width=0.50\linewidth]{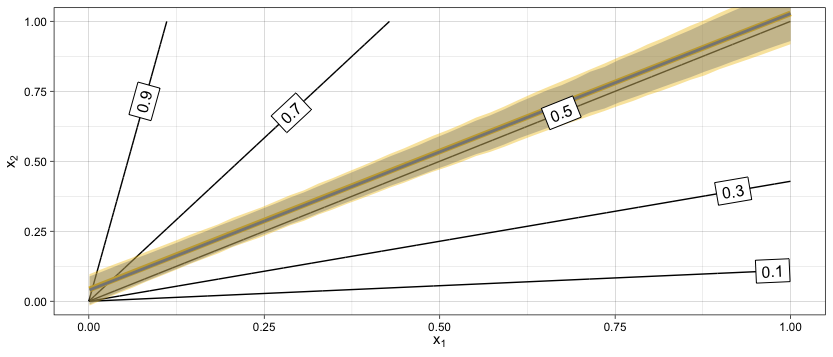}\label{fig:b_fig1_sim1}}
		\caption{Optimal decision boundaries (black lines) for target thresholds 0.1, 0.3, 0.5, 0.7, 0.9. Posterior mean boundaries for SB (grey) and TB (yellow) when targeting the (a) 0.3, and (b) 0.5  boundary. Shaded regions represent 90\% highest predictive density (HPD) regions.} 
		\label{fig1_sim1}
    \end{centering}
\end{figure}

We first evaluate the performance of tailoring by extending a simulation from \cite{hand2003local}. We simulate $n$ data points according to two covariates, $x_1$ and $x_2$, and assign label 1 with probability: $\theta := p(y = 1| x_1, x_2) = \frac{q x_2}{x_1 + q x_2}$ with $y \sim Bernoulli(\theta)$, $x_1, x_2 \sim \mathcal{U}(0, 1)$ and where $q$ is a scalar. The parameter $q$ determines the relative prevalence of the two classes, when $q > 1$ there are more class 1 than class 0, otherwise there are more class 0 than class 1. Figure \ref{fig1_sim1} shows the optimal decision boundaries in the covariate space for a range of target thresholds using $n=5000$ and $q=1$ (which leads to a prevalence of 0.5). A key feature is that these boundaries are linear, but not parallel. The absence of parallelism renders any linear  model  unsuitable  as  a  global  fit,  but  the  linearity  of  the  decision boundaries allows linear models to describe these boundaries sufficiently.

We use the decision boundaries corresponding to 0.3 and 0.5 target thresholds as exemplars. \SB results in a sub-optimally estimated decision boundary for $t=0.3$ (Figure \ref{fig:a_fig1_sim1}). The estimated 0.3 boundary from \SB is parallel to the 0.5-optimal boundary. This is expected because under this simulation setting logistic regression is bound to find a compromise model which should be linear with level lines roughly parallel to the true 0.5 boundary (where misclassification costs are equal). On the other hand, \TB allows derivation of a decision boundary which is far closer to the optimum. Note the wider predictive regions of tailoring. This is an expected consequence of our framework which we comment on in the \supple Section S9. When deriving decision boundaries under the equal costs implied by a 0.5 target threshold (Figure \ref{fig:b_fig1_sim1}), the two models are almost indistinguishable. 

\begin{figure}[H]
	\begin{centering}
	\includegraphics[width = 0.8\textwidth]{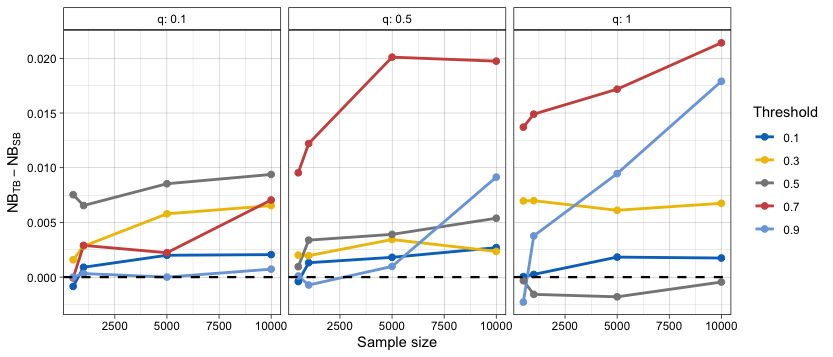}
    \caption{Difference in Net Benefit for samples sizes of 500, 1000, 5000, 10000 averaged over 20 repetitions. A positive difference means \TB outperforms \SB. The values of 0.1, 0.5, 1, for the $q$ parameter correspond to prevalence of around 0.15, 0.36, 0.50, respectively.} %The target thresholds are 0.1, 0.3, 0.5, 0.7, 0.9.}
    \label{fig2_sim1}
    \end{centering}
\end{figure}

To systematically investigate the performance of tailoring across a wide range of settings, we set-up different scenarios by varying: (1) the sample size, (2) the prevalence of the outcome, (3) and the target threshold. Model performance is evaluated in an independently sampled dataset of size 2000. Under most scenarios tailoring outperforms standard Bayesian regression (Figure \ref{fig2_sim1}). The performance gains are evident even for small sample sizes. With a few exceptions (most notably $t = 0.7$ and 0.9) the advantage of tailoring is relatively stable across sample sizes. The advantage of tailoring persists even when varying the prevalence of the outcome. In fact, we see that under certain scenarios \TB is superior to \SB even for the 0.5 boundary. Figure S2 (see \supple) illustrates such a scenario for $q = 0.1$, which corresponds to prevalence of 0.15. Under such class imbalance, which is common in medical applications, even when targeting the 0.5 boundary, one might want to use tailoring over standard modelling approaches. 

%%%%%%%%%%%%%%%%%%%%%%%%%%%%%%%%%%%%%%%%%%%%%%%%%%%%%%%%%%%%%%%%%%%%%%%%%%%%%%%%%%
\subsection{Simulation 2: Quadratic Decision Boundaries}
\begin{figure}[H]
	\begin{centering}
		\subfloat[a][]{\includegraphics[width=0.50\linewidth]{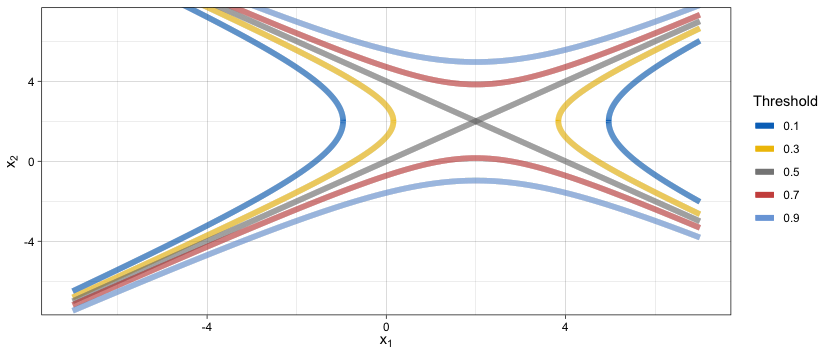}\label{fig:a_sim2}} \\
		\subfloat[b][]{\includegraphics[width=0.50\linewidth]{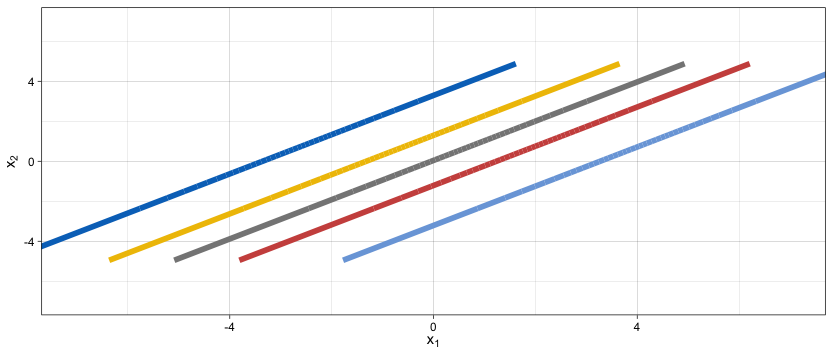}\label{fig:b_sim2}}
		\subfloat[c][]{\includegraphics[width=0.50\linewidth]{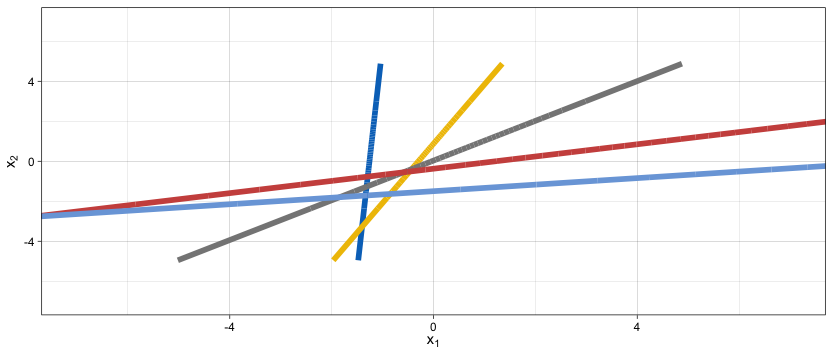}\label{fig:c_sim2}}
		\caption{(a) Optimal decision boundaries for target thresholds 0.1, 0.3, 0.5, 0.7, 0.9. Posterior median boundaries for (b) \SB, and (c) \TB.} 
		\label{fig1_sim2}
    \end{centering}
\end{figure}

Our second simulation is a more pragmatic scenario where the optimal decision boundaries are a quadratic rather than a linear function of the covariates. The model is of the form
\begin{eqnarray*}
        \mathbf{x} | y = 1
        & \sim & \mathcal{N} \left(\left[\begin{array}{c}
        1\\
        0\\
        \end{array}\right], \left[ \begin{array}{cc}
        1, 0 \\
        0, 2 
    \end{array}\right] \right)
\end{eqnarray*}
\begin{eqnarray*}
        \mathbf{x} | y = 0
        & \sim & \mathcal{N} \left(\left[\begin{array}{c}
        0\\
        1\\
        \end{array}\right],\left[ \begin{array}{cc}
        2, 0 \\
        0, 1 		 
        \end{array} \right] \right)
\end{eqnarray*}

where $\mathbf{x} = (x_1, x_2)^T$ contains the two continuous-valued predictors. The marginal probabilities of the outcome are equal, \ie $p(y = 0) = p(y = 1) = 0.5$. In this case of unequal covariance matrices, the optimal decision boundaries are a quadratic function of $\mathbf{x}$ (Figure \ref{fig:a_sim2}) (\cite{duda2012pattern}, Chapter 2). A linear model, like the one we implement is sub-optimal. Nevertheless, this example allows us to demonstrate in an analytically tractable way the advantage of tailoring and it allows us to explore a broader array of generic simulation examples, since arbitrary Gaussian distributions lead to decision boundaries that are general hyperquadrics.

Figure \ref{fig:b_sim2} and \ref{fig:c_sim2} shows the posterior median decision boundaries for \SB and \TB using $n = 5000$ under the data generating model described above, and for a range of target thresholds. It is clear that the direction of the optimal decision boundary is a function of the costs. The parallel decision boundaries obtained by applying different thresholds to the standard logistic predictions are clearly not an optimal solution when comparing against the optimal boundaries depicted in Figure \ref{fig:a_sim2}. Although limited to estimation of linear boundaries, tailoring is able to adapt the angle of the boundary to better approximate the optimal curves. One exception in comparative performance is the 0.5 threshold which is estimated perfectly for both models. This is expected, since the standard logistic model targets the 0.5 boundary. 

\begin{figure}[H]
	\begin{centering}
	\includegraphics[width = 0.8\textwidth]{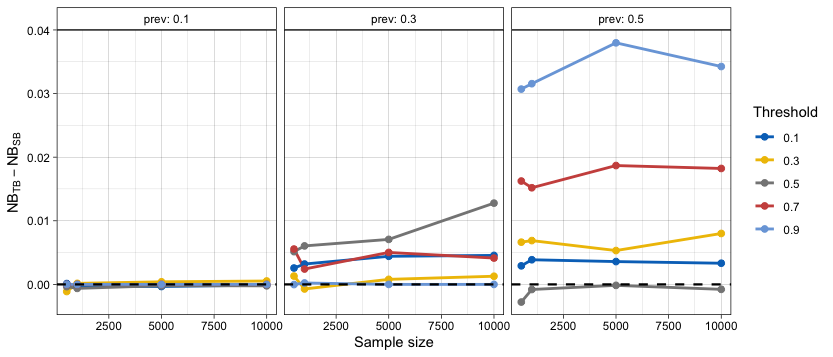}
    \caption{Difference in Net Benefit for samples sizes of 500, 1000, 5000, 10000 averaged over 20 repetitions. A positive difference means \TB outperforms \SB. Each grid corresponds to a different prevalence setting.}%The target thresholds are 0.1, 0.3, 0.5, 0.7, 0.9.}
    \label{fig4_sim2}
    \end{centering}
\end{figure}

As before, we investigate the performance of tailoring across a wide range of settings, by varying: (1) the sample size, (2) the prevalence of the outcome, (3) and the target threshold.  Performance is evaluated in an independently sampled test set of size 2000.
Figure \ref{fig4_sim2} shows the difference in $NB$ between \TB and \SB. Tailoring performs similarly or better than standard regression across all target thresholds for prevalence scenarios 0.3 and 0.5. For 0.1 the two models are closely matched. A further comparison with a non-linear model, namely Bayesian Additive Regression Trees (BART) \citep{bartpackage} is detailed in the \supple (Section S7). Briefly, \TB demonstrated equivalent or better performance than BART at the clinically relevant lower disease prevalences of 0.1 and 0.3, indicating that the benefits offered by TB cannot be matched simply by switching to a non-linear modelling framework.

%%%%%%%%%%%%%%%%%%%%%%%%%%%%%%%%%%%%%%%%%%%%%%%%%%%%%%%%%%%%%%%%%%%%%%%%%%%%%%%%%%
\subsection{Simulation 3: Data contamination}\label{sim3_contamination}
\begin{figure}[H]
	\begin{centering}
	\includegraphics[width = 0.8\textwidth]{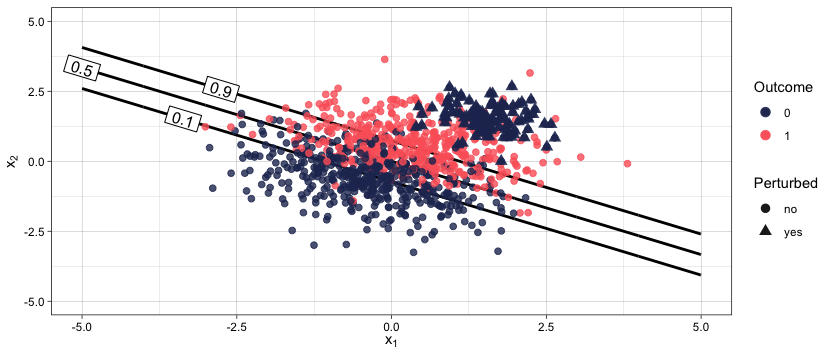}
    \caption{Single realisation from contaminated distribution with $10\%$ corrupted datapoints. Data ($n = 1000$) with labels 0 and 1 are shown in blue and red, respectively. The corrupted data points are depicted with triangles on the upper right-hand corner of the data cloud. The lines corresponds to target thresholds 0.1, 0.5, and 0.9.}
    \label{fig1_sim3}
    \end{centering}
\end{figure}

Our third simulation scenario demonstrates the robustness of tailoring to data contamination \ie the situation in which a fraction of the data have been mislabelled. The data generating model is a logistic regression with a large fraction of mislabelled datapoints. We simulate $d = 2$ covariates and $n = 1000$ datapoints.  
Figure \ref{fig1_sim3} depicts a scenario with $10\%$ of datapoints mislabelled 
among those with high values of both covariates, \ie among the upper right hand side of the data cloud. For each covariate, $1000$ values are independently drawn from a standard Gaussian distribution. Denoting the coefficient vector by $\bm{\beta} \in \mathbb{R}^{3}$ with values $\bm{\beta} = (0, 2, 3)$ (the first value corresponds to the intercept term) we simulate the outcome vector as $y \sim Bernoulli \Big( \frac{\exp \{ {\mathbf{x}^T \bm{\beta}} \}}{1 + \exp\{ {\mathbf{x}^T \bm{\beta}}\}} \Big)$, where $\mathbf{x} = (1, x_1, x_2)^T$. We then corrupt the data with class 0 datapoints, \ie we set $y := 0$ for $\psi n$ datapoints where $\psi$ is the fraction of contamination taking values $5\%, 10\%, 15\%, 20\%$ and $30\%$. The covariates are generated from equivalent and independent normal distributions, specifically $x_1, x_2 \sim \mathcal{N}(1.5, 0.5)$. This type of contamination framework has been popularised by \cite{huber1964, huber1965} and used extensively to study the robustness of learning algorithms to adversarial attacks in general \citep{balakrishnan2017computationally, diakonikolas2018sever, prasad2018robust, osama2019robust} and medical applications \citep{paschali2018generalizability}. 

\begin{figure}[H]
	\begin{centering}
	\includegraphics[width = 0.8\textwidth]{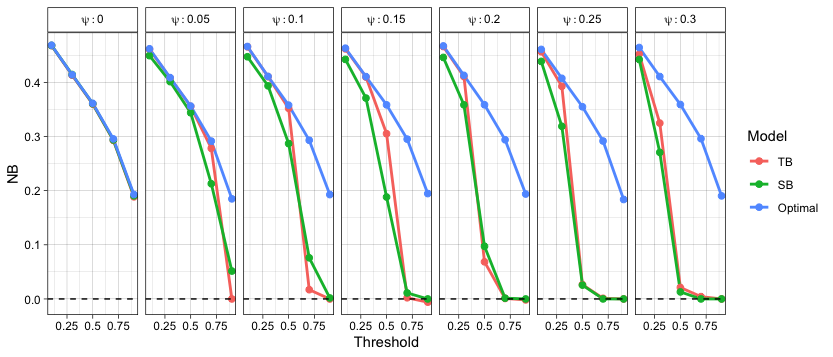}
    \caption{Net Benefit of tailoring (red) and standard regression (green) compared to optimal classification (blue) averaged over 20 repetitions. Each grid corresponds to different contamination fraction.}
    \label{fig2_sim3}
    \end{centering}
\end{figure}

We derive the optimal $NB$ based on the true probability score in an independent non-contaminated test dataset of size $n = 2000$. Figure \ref{fig2_sim3} shows the results for various contamination fractions. For most fractions \TB outperforms \SB. As the contamination fraction gets larger the performance of both models degrades, but standard regression degrades at a faster rate. Tailoring can accommodate various degrees of contamination better than standard regression, while generally never resulting in poorer performance.

Note that under no contamination (\ie $\psi = 0$, first panel Figure \ref{fig2_sim3}) \SB is an optimal classifier, since the optimal decision boundaries are parallel straight lines (Figure \ref{fig1_sim3}). However, for all other scenarios even a data corruption as small as 5\%  results in poor performance under \SB for target thresholds $>0.5$. On the contrary, tailoring maintains stable performance and close to the optimal for $t< 0.5$, for up to 15\% of mislabelled datapoints. 

%%%%%%%%%%%%%%%%%%%%%%%%%%%%%%%%%%%%%%%%%%%%%%%%%%%%%%%%%%%%%%%%%%%%%%%%%%%%%%%%%%
%%%%%%%%%%%%%%%%%%%%%%%%%%%%%%%%%%%%%%%%%%%%%%%%%%%%%%%%%%%%%%%%%%%%%%%%%%%%%%%%%%
%%%%%%%%%%%%%%%%%%%%%%%%%%%%%%%%%%%%%%%%%%%%%%%%%%%%%%%%%%%%%%%%%%%%%%%%%%%%%%%%%%
\section{Real data applications}\label{real_data}
We evaluate the performance of \TB on three real-data applications involving a breast cancer prognostication task (Section \ref{Real_data_application_1}), a cardiac surgery prognostication task (Section \ref{Real_data_application_2}) and a breast cancer tumour classification task (\supple Section S8). Overall, our empirical results demonstrate the improvement in predictive performance when taking into consideration misclassification costs during model training. 

%%%%%%%%%%%%%%%%%%%%%%%%%%%%%%%%%%%%%%%%%%%%%%%%%%%%%%%%%%%%%%%%%%%%%%%%%%%%%%%%%%
\subsection{Real data application 1: Breast cancer prognostication}\label{Real_data_application_1}
Here, we apply the \TB methodology to predict mortality after diagnosis with invasive breast cancer. The training data is based on 4718 oestrogen receptor positive subjects diagnosed in East Anglia, UK between 1999 and 2003. The outcome modelled is 10-year mortality. The covariates are age at diagnosis, tumor grade, tumor size, number of positive lymph nodes, presentation (screening vs. clinical), and type of adjuvant therapy (chemotherapy, endocrine therapy, or both). We use 20\% of the data as design and the rest as development set (see Figure S1, \supple), repeating the design/development set split $m = 5$ times. The entire train data is used to fit \SB. Both models are evaluated in an independent test set consisting of 3810 subjects. Detailed information on the datasets can be found in \cite{karapanagiotis2018development}.

An important part of the \TB methodology is the choice of $t$. In breast cancer, accurate predictions are decisive because they guide treatment. In clinical practice, treatment is given if it is expected to reduce the predicted risk by at least some pre-specified magnitude. For instance, clinicians in the Cambridge Breast Unit (Addenbrooke's Hospital, Cambridge, UK) currently use the absolute 10-year survival benefit from chemotherapy to guide decision making for adjuvant chemotherapy as follows:  $<3\%$ no chemotherapy; $3\%-5\%$ chemotherapy discussed as a possible option; $>5\%$ chemotherapy recommended \citep{down2014effect}. Following previous work \citep{karapanagiotis2018development}, we assume that chemotherapy reduces the 10-year risk of death by 22\% \citep{peto2012comparisons}. Then, a risk reduction between 3\% and 5\%, corresponds to target thresholds between 14\% and 23\%. Hence, we explore misclassification cost ratios corresponding to $t$ in the range between 0.1 and 0.5.

\begin{figure}[H]
	\begin{centering}
	\includegraphics[width = 0.8\textwidth]{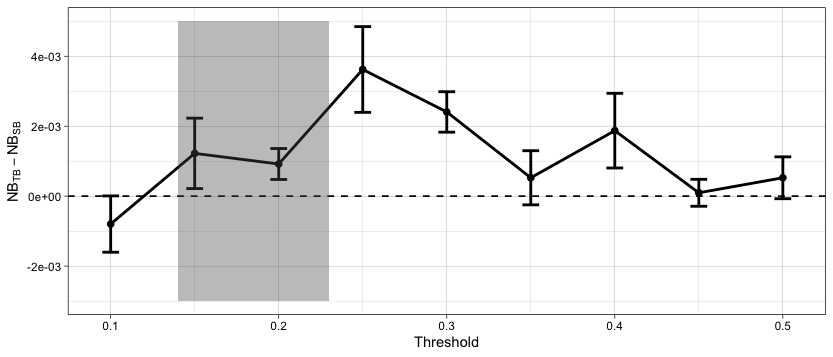}
    \caption{Difference in Net Benefit for various $t$ values evaluated on the test set. Error bars correspond to one standard error of the difference. That is, denoting the difference in Net Benefit $D^i = NB^i_{\TB} - NB^i_{\SB}$ with $i = 1, \dots, m = 5$ for each $t$ then the standard error of the difference is $SE_D = \sqrt{\frac{\sum (D^i - \bar{D})^2}{m(m-1)}}$, where $\bar{D} = \sum_i D^i / m$. This accounts for the fact that both models have been evaluated on the same data. The units on the y axis may be interpreted as the difference in benefit associated with one patient who would die without treatment and who receives therapy. The 0.14 to 0.23 shaded area on the x axis corresponds to 3\%–5\% absolute risk of death reduction with and without chemotherapy. These are the risk ranges where chemotherapy is discussed as a treatment option.}
    \label{fig_1_real_data_1}
    \end{centering}
\end{figure}

Figure \ref{fig_1_real_data_1} shows the difference in $NB$ between the two models averaged over the 5 splits. We see \TB outperforms \SB for most target thresholds, especially where decisions about adjuvant chemotherapy are made. Compared to \SB, tailoring achieves up to 3.6 more true positives per 1000 patients (when $t = 0.15$), which is equivalent to having 3.6 more true positives per 1000 patients for the same number of unnecessary treatments. %Similarly, the net benefit from using the tailored model to prescribe chemotherapy, compared with standard modelling, could be as high as 3.6 \textperthousand.

\begin{figure}[H]
	\begin{centering}
	\includegraphics[width = 0.8\textwidth]{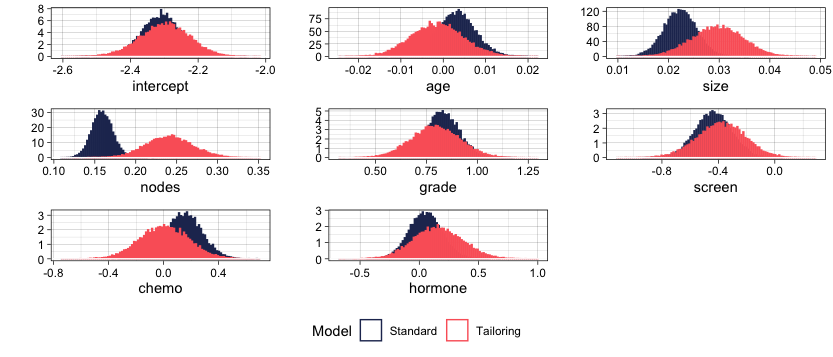}
    \caption{Marginal density plots of posterior parameters for $t = 0.15$ for \SB (blue) and \TB (red).}
    \label{fig_2_real_data_1}
    \end{centering}
\end{figure}

Next, we examine the effect of tailoring on the posterior distributions of the coefficients. As an exemplar, we use the posterior samples for the model corresponding to $t = 0.15$ (Figure \ref{fig_2_real_data_1}). We see that tailoring affects both the location and spread of the estimates compared to standard modelling. First, note the wider spread of tailoring compared to the standard models. Second, the tailored posteriors are centred on different values. The most extreme example is the coefficient for the number of nodes. Under tailoring it has a stronger positive association with the risk of death. To quantify the discrepancy between the posteriors of the two models table \ref{overlapping_dist} shows estimates of the overlapping area between the posteriors for each covariate. These range from 3\% to 70\%. The relative shifts in magnitude of the effect sizes indicates different relative importance of the covariates in terms of their contribution to the predictions from the two models. 

\begin{table}[ht]
\centering
\begin{tabular}{lr}
  \hline
  Covariate & Posterior Overlap (\%)\\ 
  \hline
  nodes & 3.05 \\ 
  size & 23.46 \\ 
  chemo & 41.92 \\ 
  age & 48.78 \\ 
  hormone & 57.76 \\ 
  grade & 62.66 \\ 
  screen & 69.94 \\ 
  %intercept & 79.04 \\ 
   \hline
\end{tabular}
\caption{Overlapping area of posterior distributions for each coefficient based Gaussian kernel density estimations \citep{pastore2019measuring}.}
\label{overlapping_dist}
\end{table}

%%%%%%%%%%%%%%%%%%%%%%%%%%%%%%%%%%%%%%%%%%%%%%%%%%%%%%%%%%%%%%%%%%%%%%%%%%%%%%%%%%
\subsection{Real data application 2: Cardiac surgery prognostication}\label{Real_data_application_2}
For our second case study we investigate whether \TB allows for better predictions, and consequently improved clinical decisions for patients undergoing aortic valve replacement (AVR). Cardiac patients with severe symptomatic aortic stenosis are considered for surgical AVR (SAVR). Given that SAVR is typically a high-risk procedure, transcatheter aortic valve implantation (TAVI) is recommended as a lower risk alternative but it is associated with higher rates of complications \citep{baumgartner2017}. The European System for Cardiac Operative Risk Evaluation (EuroSCORE) is routinely used as a criterion to choose between SAVR and TAVI \citep{roques2003logistic}. EuroSCORE is an operative mortality risk prediction model which takes into account 17 covariates encompassing patient-related, cardiac and operation-related characteristics. It was first introduced by \cite{nashef1999european} and it has been updated in 2003 \citep{roques2003logistic} and 2012 \citep{nashef2012euroscore}. Published guidelines recommend TAVI over SAVR if a patient's predicted mortality risk is above 10\% \citep{baumgartner2017} or 20\% \citep{vahanian2008transcatheter}. Here, we compare the performance of \TB with EuroSCORE and \SB given these target thresholds.

\begin{figure}[H]
	\begin{centering}
	\includegraphics[width=0.8\textwidth]{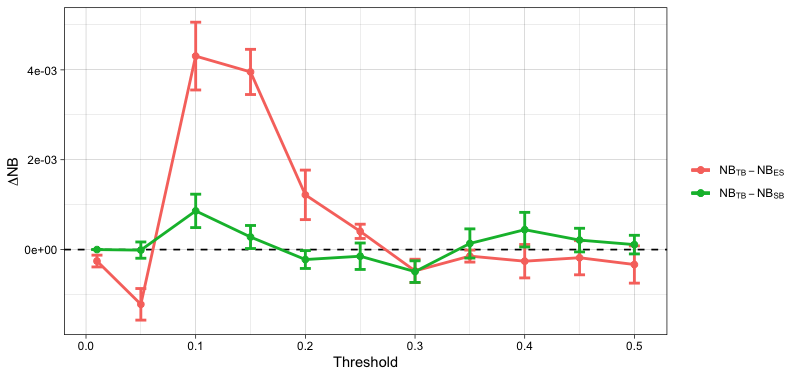}
    \caption{Difference in Net Benefit ($\Delta$NB) between \TB and EuroSCORE (ES) (red), and between \TB and \SB (green) for various target thresholds evaluated on the test set. Error bars correspond to one standard error of the difference (see caption of Figure \ref{fig_1_real_data_1} for details).}
    \label{fig1_real_data_2}
    \end{centering}
\end{figure}

We use data ($n$ = 9031) from the National Adult Cardiac Surgery Audit (UK) collected between 2011 and 2018. We use 80\% of the data for training and the rest for testing, repeating the train/test set split $m = 5$ times. For this data a design set to estimate $\pi_u(\mathbf{x}_i)$ is not necessary (see \supple Figure S1) but instead we use the predictions from EuroSCORE \citep{roques2003logistic}. We add an extra step of re-calibration to account for the population/time drift \citep{cox1958two, miller1993validation}. Figure \ref{fig1_real_data_2} presents the results. We see \TB outperforms both EuroSCORE and \SB when targeting the 0.1 threshold, and only EuroSCORE at $t = 0.2$. 

\begin{figure}[H]
	\begin{centering}
	\includegraphics[width=0.8\textwidth]{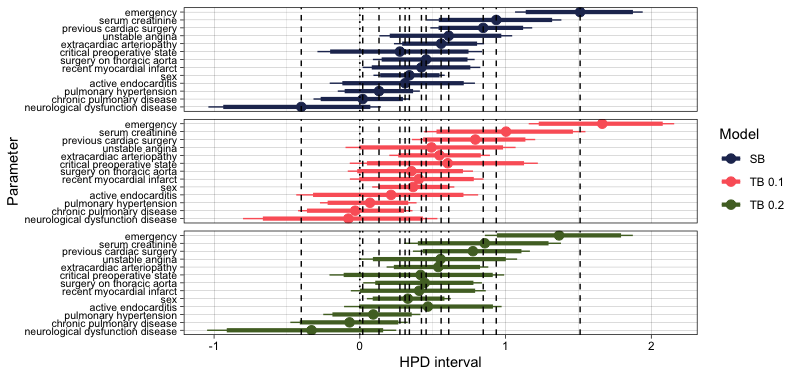}
    \caption{Highest posterior density (HPD) regions for the parameters. Dots represent medians, and thick and thin lines represent 90 and the 95\% of the HPD regions, respectively. The dashed vertical lines pass through the posterior median values of the \SB parameters.}
    \label{Aortic_HPD_intervals}
    \end{centering}
\end{figure}

We further investigate the effect of tailoring to individual parameters. Figure \ref{Aortic_HPD_intervals} shows the highest posterior density (HPD) regions for a subset of the covariates under \SB and \TB for $t= 0.1$ and $0.2$. As in the previous case study, under tailoring the regions are generally wider and are centred on different values. For instance, compared to \SB under both $t = 0.1$ and 0.2 the posteriors of critical operative state and unstable angina are shifted towards the same direction (positive for critical operative state and negative for unstable angina). Contrast these with the posterior of emergency that compared to \SB it is centred on more positive values under $t =0.1$ and more negative under $t =0.2$. On the contrary, extracardiac arteriopathy, recent myocardial infarct, and sex are centred on similar values across the three models.  This once more exemplifies the change in the contribution of some covariates towards the predicted risks when taking into account misclassification costs. 

%%%%%%%%%%%%%%%%%%%%%%%%%%%%%%%%%%%%%%%%%%%%%%%%%%%%%%%%%%%%%%%%%%%%%%%%%%%%%%%%%%
%%%%%%%%%%%%%%%%%%%%%%%%%%%%%%%%%%%%%%%%%%%%%%%%%%%%%%%%%%%%%%%%%%%%%%%%%%%%%%%%%%
%%%%%%%%%%%%%%%%%%%%%%%%%%%%%%%%%%%%%%%%%%%%%%%%%%%%%%%%%%%%%%%%%%%%%%%%%%%%%%%%%%
\section{Discussion}\label{discussion}
In this work, we present Tailored Bayes, a framework to incorporate misclassification costs into Bayesian modelling. We demonstrate that our framework improves predictive performance compared to standard Bayesian modelling over a wide range of scenarios  in which the costs of different classification errors are unbalanced. 

%An attractive feature of the methodology is its wide applicabilty. The tailored posterior integrates the attractive features of Bayesian inference - such as flexible hierarchical modelling and the use of prior information - albeit allowing for targeted inference. Additionally, TB is flexible. 

The methodology relies solely on the construction of the datapoint-specific weights (see (\ref{weights})). In particular, we need to specify $t$, the grid of $\lambda$ values for the CV, a model to estimate $\pi_u(\mathbf{x}_i)$ and the weighting function, $h$. For some applications there may be a recommended target threshold, $t$. For instance, UK national guidelines recommend that clinicians use a risk prediction model (QRISK2; \citep{hippisley2008predicting}) to determine whether to prescribe statins for primary prevention of cardiovascular disease (CVD) if a person's CVD risk is 10\% or more \citep{nice2016cardiovascular}. When guidelines are not available, the specification of $t$ is inevitably subjective, since it reflects the decision maker's preferences regarding the relative costs of different classification errors.  In practice, eliciting these preferences may be challenging, despite the numerous techniques that have been proposed in the literature to help with this (\eg \cite{tsalatsanis2010regret, hunink2014decision}). In such situations, we advocate fitting the model for a range of plausible $t$ values that reflect general decision preferences.  For example, research in both mammographic \citep{schwartz2000us} and colorectal cancer screening \citep{boone2013patients} has shown that healthcare professionals and patients alike greatly value gains in sensitivity over loss of specificity. For additional examples on setting $t$ see \cite{vickers2016net} and \cite{wynants2019three}. Further examples in which benefits and costs associated with an intervention (as well as with patients' preferences) are taken into account, are provided by \cite{manchanda2016specifying, le2017decision, watson2020evaluating}.
 
We discuss the remaining elements for the construction of the weights in the supplementary material (Section S9). There we define the effective sample size for tailoring, $ESS_t$, and showcase how to use it to set the upper limit for the grid of $\lambda$ values. In addition, we show our framework is robust to miscalibration of $\pi_u(\mathbf{x}_i)$ and the choice of $h$. The framework is therefore flexible, allowing many ways for the user to specify the weights. 

%More specifically, we need to specify $t$, the grid of $\lambda$ values for the CV, a model to estimate $\pi_u(\mathbf{x}_i)$ and the weighting function, $h$. In the \supple (Section S7) we present a few considerations on the choice of $t$. We further define the effective sample size for tailoring, $ESS_t$, and showcase how to use it to set the  upper limit for the grid of $\lambda$ values. In addition, we show our framework is robust to miscalibration of $\hat{\pi}_u(\mathbf{x}_i)$ and the choice of $h$. The framework is therefore flexible, allowing many ways for the user to specify the weights.

%Another attractive feature of the methodology is its wide applicability. 
%Our framework is built upon the seminal work of \cite{hand2003local}. 
In contrast to the work of \cite{hand2003local} our approach is framed within the Bayesian formalism. Consequently, the tailored posterior integrates the attractive features of Bayesian inference - such as flexible hierarchical modelling, the use of prior information and quantification of uncertainty - while also allowing for tailored inference. Quantification of uncertainty is critically important, especially in healthcare applications  \citep{begoli2019need, kompa4second}. Whilst two (or more) models can perform similarly in terms of aggregate metrics (\eg area under ROC curve) they can provide very different individual (risk) predictions for the same patient \citep{pate2019uncertainty, li2020consistency}. This can ultimately lead to different decisions for the individual, with potential detrimental effects. Uncertainty quantification can mitigate this issue since it allows the clinician to abstain from utilising the model’s predictions. If there is high predictive uncertainty for an individual, the clinician can discount or even disregard the prediction. 

\begin{figure}[H]
	\begin{centering}
	\includegraphics[width = 0.8\textwidth]{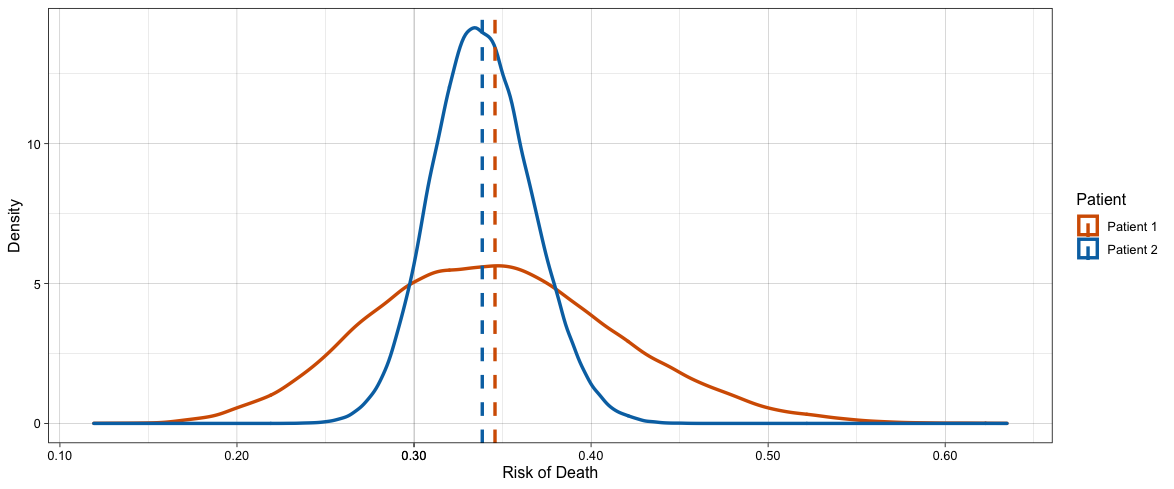}
    \caption{Predictive uncertainty for the risk of death in two patients. These posterior predictive distributions reflect the range of risks assigned to these patients, and the mean risk is shown as vertical lines. Despite the fact that both patients have similar mean risks, we may be more inclined to trust the predictions for patient 2 given the lower amount of uncertainty associated with that prediction.}
    \label{indiv_predictive_uncertainty}
    \end{centering}
\end{figure}

To illustrate this point, we use the \SB posterior from the breast cancer prognostication case study. The posterior predictive distributions for two patients are displayed in Figure \ref{indiv_predictive_uncertainty}. The average posterior risk for each patient is indicated by the vertical line at 34 and 35\%, respectively. Based solely on these average estimates chemotherapy should be recommended as a treatment option to both patients (see Section \ref{Real_data_application_1}). It is clear, however, that the predictive uncertainty for these two patients is quite different, as the distribution of risk for patient 1 is much more dispersed than the distribution for patient 2. One way to quantify the predictive uncertainty would be to calculate the standard deviation of these distributions, which are 6.9\% and 2.8\% for patient 1 and patient 2, respectively. Even though both estimates are centred at similar values the predictive uncertainty for patient 1 is more than two times higher than patient 2. Using this information, we could flag patient 1 as needing more information before making a clinical decision.  

%In addition, for models that predict critical conditions (\eg sepsis), uncertainty estimates is vital for triaging patients. Clinicians could focus on patients with highly certain model estimates, but also further examine patients for whom the model is uncertain with respect to their current condition. For patients with highly uncertain predictions, additional lab values or more frequent monitoring could be requested.

%Consequently, the tailored posterior integrates the attractive features of Bayesian inference - such as flexible hierarchical modelling and the use of prior information - while also allowing for tailored inference. 
A few related comments are in order.
In this work we use vague Gaussian priors, but they could be replaced with other application-specific distribution choices. For instance, in the case of high-dimensional data another option could be the sparsity-inducing prior used by Bayesian lasso regression \citep{park2008bayesian}. Furthermore, we can easily incorporate external information in a flexible manner, through $\pi_u(\mathbf{x})$, in addition to the prior on the coefficients. %Incorporation of external information (\eg from historical data) is of interest given the proliferation of risk prediction models. 
If a well-established  model exists then it is  natural to consider using it to improve the performance of an expanded model. We have implemented such an approach in Section \ref{Real_data_application_2}. \cite{cheng2019informing} propose several approaches for incorporating published summary associations as prior information when building risk models. A limitation of their approaches is the requirement for a parametric model, \ie information on regression coefficients. Our method does not have any restriction on the form of $\pi_u(\mathbf{x})$, it can arise from a parametric or non-parametric model. 

We note that we opted to use the same set of covariates, $\mathbf{x}$, to estimate both $\pi_{w_i}(\mathbf{x};\bm{\beta})$ and $\pi_u(\mathbf{x})$. This does not need to be the case. If available, we could instead use another set of covariates, say $\mathbf{Z}$ to estimate $\pi_u(\mathbf{z})$. The set $\mathbf{Z}$ could be a superset or a subset of $\mathbf{X}$ or the two sets could be completely disjoint. We also note that in this work we focus on linear logistic regression to showcase the methodology (linear refers to linear combinations of the covariates). This is because it is widely utilised and allows analytical and computational tractability. Nevertheless, we would stress that our framework is generic, and not restricted to linear logistic regression. It can accommodate a wide range of modelling frameworks, from linear to non-linear and from classical statistical approaches to state-of-the-art machine learning algorithms. As a result, future work could consider such extensions to non-linear models. Also, future work could consider the advantages of a joint estimation, \ie  both steps, stage 1 (estimation of weights) and stage 2 (estimation of weighted prediction probabilities) jointly. A further direction is the extension of the framework to high-dimensional settings.

To conclude, in response to recent calls for building clinically useful models \citep{chatterjee2016developing,shah2019making} we present an overarching Bayesian learning framework for binary classification where we take into account the different benefits/costs associated with correct and incorrect classifications. The framework requires the modellers to first think of how the model will be used and the consequences of decisions arising from its use - which we would argue should be a prerequisite for any modelling task. Instead of fitting a global, agnostic model and then deploying the result in a clinical setting we propose a Bayesian framework to build towards models tailored to the clinical application under consideration. 

%
%There are calls for building clinically useful models taking into account what matters in clinical care. 
%can support the preferential use of treatments in patients
%who most benefit, while avoiding use in patients who do not benefit or might even be harmed. 
%The framework presented here urges to first think of how the model will be used and the consequences of decisions arising from its use. 
%Instead, of fitting a global, agnostic model and then deploying the result in a clinical setting we present a Bayesian framework to build towards clinically useful models. 
%context of the subsequent actions it triggers is necessary to

%%%%%%%%%%%%%%%%%%%%%%%%%%%%%%%%%%%%%%%%%%%%%%%%%%%%%%%%%%%%%%%%%%%%%%%%%%%%%%%%%%
%%%%%%%%%%%%%%%%%%%%%%%%%%%%%%%%%%%%%%%%%%%%%%%%%%%%%%%%%%%%%%%%%%%%%%%%%%%%%%%%%%
%%%%%%%%%%%%%%%%%%%%%%%%%%%%%%%%%%%%%%%%%%%%%%%%%%%%%%%%%%%%%%%%%%%%%%%%%%%%%%%%%%
\section{Software}\label{software}

The R code used for the experiments in this paper has been made available as an R package, TailoredBayes, on Github: \url{https://github.com/solonkarapa/TailoredBayes}

%\section{Supplementary Material}
%\label{sec6}

%Supplementary material is available online at
% \href{http://biostatistics.oxfordjournals.org}%
% {http://biostatistics.oxfordjournals.org}.
%\url{http://biostatistics.oxfordjournals.org}.

\section*{Acknowledgments}
The authors thank Paul Pharoah for providing the breast cancer dataset and Jeremias Knoblauch for the insightful discussions.
%Funding for the project was provided by ...
This work was supported by the Medical Research Council (MC\_UU\_00002/9 to S.K and P.J.N, MC\_UU\_00002/13 \& MR/R014019/1 to P.D.W.K). UB is supported by National Institute for Health Research Bristol Biomedical Research Centre (NIHR Bristol BRC). This work was also supported by the National Institute for Health Research (Cambridge Biomedical Research Centre at the Cambridge University Hospitals NHS Foundation Trust). The views expressed are those of the authors and not necessarily those of the NHS, the NIHR or the Department of Health and Social Care. S.K was also supported by The Alan Turing Institute under the EPSRC grant EP/N510129/1. Partly funded by the RESCUER project. RESCUER has received funding from the European Union's Horizon 2020 research and innovation programme under grant agreement No. 847912.

{\it Conflict of Interest}: None declared.

% References
\bibliographystyle{abbrvnat}
\bibliography{refs}

%%%%%%%%%%%%%%%%%%%%%%%%%%%%%%%%%%%%%%%%%%%%%%%%%%%%%%%%%%%%%%%%%%%%%%%%%%%%%%%%%%
%%%%%%%%%%%%%%%%%%%%%%%%%%%%%%%%%%%%%%%%%%%%%%%%%%%%%%%%%%%%%%%%%%%%%%%%%%%%%%%%%%
%%%%%%%%%%%%%%%%%%%%%%%%%%%%%%%%%%%%%%%%%%%%%%%%%%%%%%%%%%%%%%%%%%%%%%%%%%%%%%%%%%
% supplementary
\newpage

%\usepackage[ruled,vlined]{algorithm2e}
%\include{pythonlisting}
%\usepackage{algpseudocode}

% commands for S -sections 
\setcounter{equation}{0}
\setcounter{figure}{0}
\setcounter{table}{0}
\setcounter{section}{0}
\renewcommand{\theequation}{S\arabic{equation}}
\renewcommand{\thefigure}{S\arabic{figure}}
\renewcommand{\bibnumfmt}[1]{[S#1]}
\renewcommand{\citenumfont}[1]{S#1}
\renewcommand{\thesection}{S\arabic{section}}  
\renewcommand{\thetable}{S\arabic{table}} 

\SetAlgorithmName{Pseudo-code}{}{} % change name
\SetKwInput{KwInput}{input}  % Set the Input
\SetKwInput{KwOutput}{output}

% Title
%\title{Supplementary Materials for \say{Tailored Bayes: a risk modeling framework under unequal misclassification costs}}

%\beginsupplement
%\documentclass{bio} % to use
%\documentclass[a4paper]{article}
%\documentclass{article}
%\documentclass[class=article, crop=false]{standalone}
%\usepackage[subpreambles=true]{standalone}
%\usepackage{import}
%\usepackage{blindtext}

%\usepackage{dirtytalk}
%\usepackage{bm}
%\usepackage{xspace} % removes extraneous spaces after user-defined commands
%\usepackage{float}
%\usepackage{subfig}
%\usepackage{amsthm,multirow,amsmath,fancybox,amssymb}

% for revisions
%\usepackage{changes}

%% Sets page size and margins
%\usepackage[a4paper,left = 1in, right = 1in]{geometry}

% commands for S -sections 

%%%%%%%%%%%% custom  %%%%%
%% math symbols

%\begin{document}

\textbf{\LARGE Supplementary Materials for \say{Tailored Bayes: a risk modelling framework under unequal misclassification costs}}\\

% List of authors, with corresponding author marked by asterisk
\author{Solon Karapanagiotis, Umberto Benedetto, Sach Mukherjee, Paul D. W. Kirk, Paul J. Newcombe\\[4pt]}

% Running headers of paper:
%\markboth%
% First field is the short list of authors
%{S. Karapanagiotis and others}
% Second field is the short title of the paper
%{Tailored Bayes for risk prediction}

%\maketitle

%%%%%%%%%%%%%%%%%%%%%%%%%%%%%%%%%%%%%%%%%%%%%%%%%%%%%%%%%%%%%%%%%%%%%%%%%%%%%%%%%%
\section{Interpretation of the TB prior (and posterior)}

Here we show the prior in \TB can be interpreted as a regularizer on a per-datapoint influence/importance. First, we slightly modify notation and consider data $D_i=  (X_i,Y_i)$ as a copy of a random variable $D = (X, Y) \in \mathbb{R}^d \times \{0, 1\}$. Then, let $\mathcal{L}(D_i|\bm{\beta})$ be the standard likelihood contribution of datapoint $i$ ($i = 1, \ldots, n)$. The \TB posterior, up to a normalising constant, is 
\begin{equation}
  p(\bm{\beta}|D) \propto \prod_{i=1}^n \mathcal{L}(D_i|\bm{\beta})^{w_i} p(\bm{\beta}).
  \label{tailored_posterior}
\end{equation}

Following \cite{walker2001bayesian} we can view (\ref{tailored_posterior}) as combining the original likelihood function with a \emph{data-dependent} prior that is divided by a portion of the likelihood. To see this, we first define the data-dependent prior as
$$ \frac{p(\bm{\beta})} {\prod_{i=1}^{n} \mathcal{L}(D_i|\bm{\beta})^{1-w_i}}$$
    
which corresponds to 

$$p(\bm{\beta}|D) \propto \prod_{i=1}^n \mathcal{L}(D_i|\bm{\beta}) \frac{p(\bm{\beta})} {\prod_{i=1}^{n} \mathcal{L}(D_i|\bm{\beta})^{1 - w_i}} 
= \prod_{i=1}^n \mathcal{L}(D_i|\bm{\beta})^{w_i} p(\bm{\beta}),$$

which is seen to coincide with (\ref{tailored_posterior}). This data-dependent downweighting of the prior reduces the weights of those parameter values that “track the data too closely” \citep{linero2018bayesian}.

%%%%%%%%%%%%%%%%%%%%%%%%%%%%%%%%%%%%%%%%%%%%%%%%%%%%%%%%%%%%%%%%%%%%%%%%%%%%%%%%%%
\section{Model inference and prediction}\label{model_inference_and_prediction}
To sample from the \TB posterior we use Markov Chain Monte Carlo (MCMC) (see Section \ref{computation} for details on the computational scheme). We obtain $S$ posterior samples $\{\bm{\beta}^s\}^S_{s=1}$, where $\bm{\beta}^s = (\beta^s_1, \dots, \beta^s_{d+1})$. We use the posterior samples to approximate the predictive density for test data $\mathbf{x}_*$
\begin{equation}
    \begin{split}
        p( \pi(\mathbf{x}_*) \mid \mathbf{x}_*, D) = \int p(\pi(\mathbf{x}_*) \mid \mathbf{x}_*, \bm{\beta}) p(\bm{\beta} \mid D) d\bm{\beta}\\
        \approx \frac{1}{S} \sum_{s=1}^S p(\pi(\mathbf{x}_*) \mid \mathbf{x}_*, \bm{\beta}^s).
   \end{split}
    \label{posterior_predictive}
\end{equation}

To calculate point predictions we summarise (\ref{posterior_predictive}) by using the posterior predictive mean,
\begin{equation}
    \begin{split}
        \hat{\pi}( \mathbf{x}_*)  = \int \pi(\mathbf{x}_*) p( \pi(\mathbf{x}_*) \mid \mathbf{x}_*, D) d\pi(\mathbf{x}_*),
   \end{split}
    \label{point_posterior_predictive}
\end{equation}

which is used as a plug-in into (5) in the \manu. We also use Bayesian inference for the estimation of $\pi_u(\mathbf{x})$ so $\pi(\mathbf{x}_*)$ can be conceptually replaced by $\pi_u(\mathbf{x})$ in eq (\ref{posterior_predictive}) and (\ref{point_posterior_predictive}) with the caveat that they are estimated in different subsets of the data (see Section \ref{data_splitting} for the data splitting stategy we are implementing).

%%%%%%%%%%%%%%%%%%%%%%%%%%%%%%%%%%%%%%%%%%%%%%%%%%%%%%%%%%%%%%%%%%%%%%%%%%%%%%%%%%
\section{Cross-validation to choose $\lambda$}\label{choosing_lambda}
We use stratified $K$-fold cross-validation (CV) to choose $\lambda$ in (7) in the \manu. The stratification ensures the prevalence of the outcome is the same in each fold. In $K$-fold CV, the data is partitioned into $K$ subsets $D_{(k)}$, for $k = 1, \dots, K$ and then the model is fit separately to each training set $D_{(-k)}$ thus yielding a posterior distribution $p(\bm{\beta}\mid D_{(-k)})$. When calculating the predictive performance of the model the data of the $k^{th}$ fold is used as test data.  The predictive density for $\mathbf{x}_*$, if it is in subset $k$, is
\begin{equation}
    \begin{split}
        p( \pi(\mathbf{x}_*) \mid \mathbf{x}_*, D_{(-k)}) = \int p(\pi(\mathbf{x}_*) \mid \mathbf{x}_*, \bm{\beta}) p(\bm{\beta} \mid D_{(-k)}) d\bm{\beta}\\
        \approx \frac{1}{S} \sum_{s=1}^S p(\pi(\mathbf{x}_*) \mid \mathbf{x}_*, \bm{\beta}^s).
   \end{split}
    \label{posterior_predictive_CV}
\end{equation}
and the posterior predictive expectation is $\hat{\pi}( \mathbf{x}_*)  = \int \pi(\mathbf{x}_*) p( \pi(\mathbf{x}_*) \mid \mathbf{x}_*, D_{(-k)}) d\pi(\mathbf{x}_*)$ which is used as a plug-in into (5) in the \manu to calculate the $K$-fold CV estimate of NB in the $k^{th}$ fold, $\text{NB}_{(k)}$. We choose $\lambda$ as 
\begin{equation*}
    \lambda^* = \argmaxA_\lambda \frac{1}{K}\sum_{k=1}^K \text{NB}_{(k)}.
\end{equation*}

We use $K=5$ for the all analysis. In practice, we have seen the results are insensitive to the choice of $K$. 

%%%%%%%%%%%%%%%%%%%%%%%%%%%%%%%%%%%%%%%%%%%%%%%%%%%%%%%%%%%%%%%%%%%%%%%%%%%%%%%%%%
\section{Data splitting strategy}\label{data_splitting}
\begin{figure}[H]
	\begin{centering}
	\includegraphics[width = 0.7\textwidth]{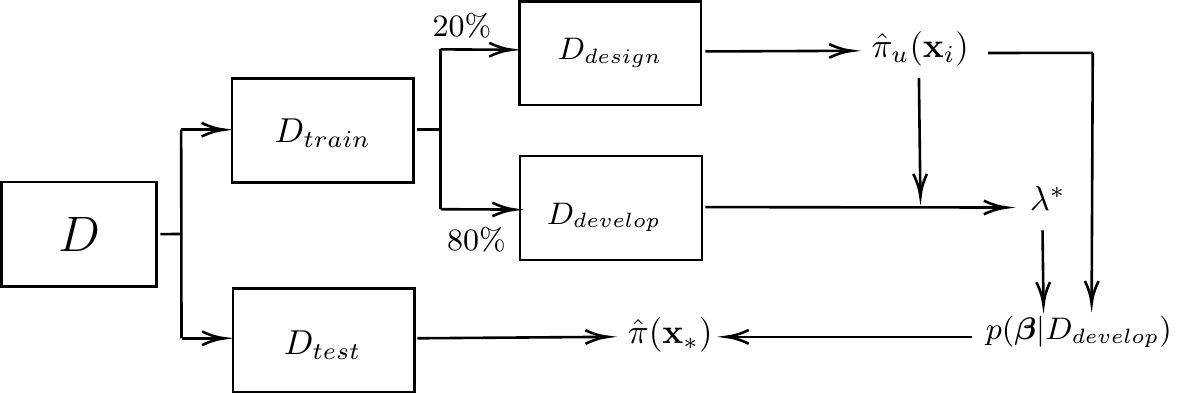}
    \caption{The data splitting strategy. The dataset, $D$, is split into train ($D_{train})$ and test ($D_{test}$) sets. The train set is subsequently split again into design ($D_{design}$) (20\%) and development ($D_{develop}$) (80\%). The design part is used to estimate $\hat{\pi}_u(\mathbf{x}_i)$. The development part is used to choose $\lambda^*$ (5-fold CV, see Section \ref{choosing_lambda}  for details). After choosing a $\lambda^*$ value the model is fit to the entire development part, obtaining the posterior, $p(\bm{\beta}|D_{develop})$. Finally, the test set is used to create predictions, $\hat{\pi}( \mathbf{x}_*) $ (this is the posterior predictive mean defined in Section \ref{model_inference_and_prediction}.)}
    \label{fig_data_splitting}
    \end{centering}
\end{figure}

To avoid overfitting due to the estimation of both $\pi_u(\mathbf{x}_i)$ and $\pi_{w_{i}}(\mathbf{x}_i;\bm{\beta})$ from the same dataset we use the following data splitting process (Figure \ref{fig_data_splitting} and Pseudo-Code \ref{pseudocode}). First, the data is split into training and testing sets. This step is avoided if we already have an independent test set (Section 4.1 \manu) or if data is simulated (Section 3 \manu). The train set is subsequently split again into design (20\%) and development (80\%). The design part is used to estimate $\pi_u(\mathbf{x}_i)$. The development part is used to choose $\lambda$ (with 5-fold stratified CV, see Section \ref{choosing_lambda} for details). After choosing a $\lambda$ value the model is fit to the entire development part, and it is with respect to the posterior from this final fit credible intervals are generated in the analyses. Finally, the test set is used to validate the performance of the model. We opted for this three-way splitting strategy, design-development-test set, because we have large datasets available but any other method such as leave-one-out, (nested) cross-validation, or bootstrap methods could be used.

%%%%%%%
%\section{The TB algorithm in pseudo-code}
   
\begin{algorithm}[H]
   
    \KwInput{$D$: dataset, 
    $K$: number of folds for the cross-validation, 
    $\lambda$: tuning parameter,
    \text{Model}$_u$: model to estimate $\pi_u(\mathbf{x}_{i})$, 
    \text{Model}$_w$: model to estimate $\pi_{w_{i}}(\mathbf{x}_{i}; \bm{\beta})$}
    
 \begin{algorithmic}
% TB procedure 
\Procedure{TB}{$D, K, \lambda, \text{Model}_u, \text{Model}_w$}
 %for $j$ in \{$1\dots m$\} do 
 
% Change the numbering style of Footnotes in LaTeX
\renewcommand{\thefootnote}{\alph{footnote}}
 
    \begin{enumerate}
        \item $D_{design}$, $D_{develop}$, $D_{test}$ $\gets$ $D$ 
        \Comment{randomly split $D$ into design, development and test sets} 
        
        \item $f_u$ $\gets$ \text{Model}$_u(D_{design})$ 
        \Comment{train a model on $D_{design}$ to estimate a function $f_u$
        \footnotemark[1]}
        
        %\item $\mathbf{x}_{develop} = \{ \mathbf{x}_1, \dots, \mathbf{x}_n \}$  \text{- write $D^j_{develop}$ as a set of $n$ datapoints}
        
        %\item $\hat{\pi}_u(\mathbf{x}_{develop}) \leftarrow \{f^j_u(\mathbf{x}_1), \dots f^j_u(\mathbf{x}_n) \}$ - apply the model $f^j$ to $\mathbf{x}_{develop}$ to obtain estimated probabilities $\hat{\pi}_u$
        
        \item $\hat{\pi}_u(\mathbf{x}_{i}) \leftarrow f_u(\mathbf{x}_i)$ \Comment{apply the function $f_u$ to $\mathbf{x}_{i}$ where $i \in D_{develop}$ to obtain $\hat{\pi}_u(\mathbf{x}_{i})$}
        \footnotemark[2]
        
        \item $\lambda^* \leftarrow \text{CV}(D_{develop}, K, \lambda$) 
        \Comment{use cross-validation to obtain $\lambda^*$} 
        \footnotemark[3]
        
        \item $f \leftarrow \text{Model}_w(D_{develop}, \lambda^*)$ 
        \Comment{train a model on $D_{develop}$ to estimate a function $f$} \footnotemark[4]
        
        %\item $\mathbf{x}_{*} = \{\mathbf{x}_1, \dots, \mathbf{x}_m \}$  \text{- write $D^j_{test}$ as a set of $m$ datapoints}
        
        \item $\hat{\pi}(\mathbf{x}_{*}) \leftarrow  f(\mathbf{x}_*)$ 
        \Comment{apply the function $f$ to $\mathbf{x}_{*} \in D_{test}$ to obtain $\hat{\pi}(\mathbf{x}_{*})$}\footnotemark[5]
        
        \EndProcedure
        
        \end{enumerate}
         
        \end{algorithmic}
         
        \KwOutput{$\hat{\pi}(\mathbf{x}_{*})$ \hspace{1.5cm}
        \Comment{the posterior predictive mean. Additionally, the posterior distribution of $f$ can be returned}}

    \vspace{0.5cm}
    
    % footnotes    
   
    \footnotetext[1]{In this work, Model$_u$ is the standard Bayesian logistic regression and $f_u(\mathbf{x}):= \mathbf{x}^T\bm{\beta}_u$. Hence, we are estimating the posterior of $\bm{\beta}_u$.}
    
    \footnotetext[2]{$\hat{\pi}_u(\mathbf{x}_{i})$ is the posterior predictive mean, which integrates over the uncertainty in $\bm{\beta}_u$, see section S2 for formal definition.}
    
    \footnotetext[3]{The CV function performs $K$-fold cross-validation using the development set as input. For the pre-specified $\lambda$ values returns $\lambda^*$, the value that gives the highest average NB, see section S3 for formal definition.}
    
    \footnotetext[4]{In this work, \text{Model}$_w$ is the tailored Bayesian logistic regression (Sections 2.3 and 2.4, \manu) and $f(\mathbf{x}):= \mathbf{x}^T\bm{\beta}$.}
    
   \footnotetext[5]{$\hat{\pi}(\mathbf{x}_{*})$ is the posterior predictive mean, which integrates over the uncertainty in $\bm{\beta}$, see section S2 for formal definition.}
    
\caption{The TB algorithm.}
\label{pseudocode}
\end{algorithm}

%%%%%%%%%%%%%%%%%%%%%%%%%%%%%%%%%%%%%%%%%%%%%%%%%%%%%%%%%%%%%%%%%%%%%%%%%%%%%%%%%%
\section{Computational scheme}\label{computation}
For all analysis in this report we use MCMC which has become a very important computational tool in Bayesian statistics since it allows for Monte Carlo approximation of complex posterior distributions where analytical or numerical integration techniques are not applicable. The Markov chain is constructed using random walk Metropolis-Hastings updates \citep{brooks2011handbook}. We give a brief overview of the algorithm.   

The target distribution is $p(\bm{\beta}| D)$ (see (10) \manu). The sampling scheme starts at an initial set of parameter values, denote these $\bm{\beta}^{0}$. To sample the next set of parameters, which we denote $\bm{\beta}^{1}$, we propose moving from the current state to another set of parameter values, $\bm{\beta}^{new}$, by using a proposal function $q(\bm{\beta}^{new}|\bm{\beta})$. We then accept these proposed values as the next sample with probability equal to the Metropolis-Hastings ratio:
\begin{equation}
    \text{MHR}(\bm{\beta}, \bm{\beta}^{new}) = \frac{L(D|\bm{\beta}^{new}) p(\bm{\beta}^{new})}{L(D|\bm{\beta}) p(\bm{\beta})} \times \frac{q(\bm{\beta}|\bm{\beta}^{new})}{q(\bm{\beta}^{new}|\bm{\beta})},
    \label{MHR}
\end{equation}

where $L(D|\bm{\beta})$ is the tailored likelihood and $p(\bm{\beta})$ the prior, given in Sections 2.4 and 2.5 of the \manu. The proposed move is accepted  with probability
\begin{equation*}
\alpha(\bm{\beta}, \bm{\beta}^{new}) = min(1, \text{MHR}(\bm{\beta}, \bm{\beta}^{new})).
\end{equation*}

If this new set of values is accepted, the proposed set is accepted as $\bm{\beta}^{1}$; otherwise, the sample value remains equal to the current sample value, \ie $\bm{\beta}^{1} = \bm{\beta}^{0}$. The proposal function is Gaussian, \ie $q \sim \mathcal{N}(\bm{\beta}, \bm{I}sd)$ where $sd$ is chosen to yield an acceptance rate $\approx 0.24$ \citep{brooks2011handbook}. In the current version of the algorithm all parameters are updated jointly. 

%%%%%%%%%%%%%%%%%%%%%%%%%%%%%%%%%%%%%%%%%%%%%%%%%%%%%%%%%%%%%%%%%%%%%%%%%%%%%%%%%%
\section{Supplementary Figures}

\begin{figure}[H]
	\begin{centering}
	\includegraphics[width = 0.7\textwidth]{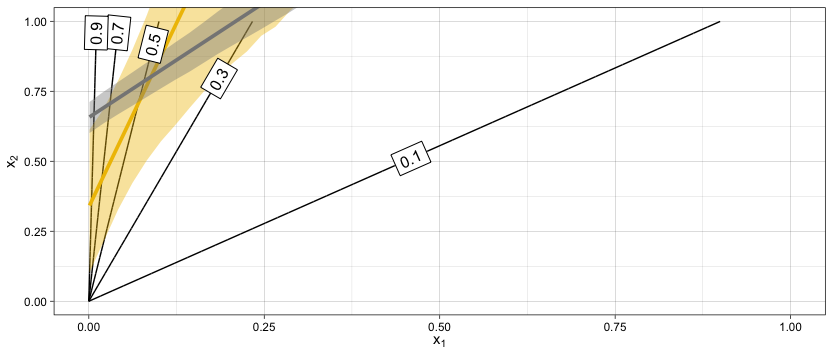}
    \caption{Single realisation from model in Section 3.1 with $q = 0.1$ corresponding to prevalence of around 0.15. Optimal decision boundaries (black lines) for target thresholds 0.1, 0.3, 0.5, 0.7, 0.9. Posterior mean boundaries for \SB (grey) and \TB (yellow) when targeting the 0.5 boundary. Shaded regions represent 90\% highest predictive density (HPD) regions.}
    \label{toy_Hand_prev}
    \end{centering}
\end{figure}

%%%%%%%%%%%%%%%%%%%%%%%%%%%%%%%%%%%%%%%%%%%%%%%%%%%%%%%%%%%%%%%%%%%%%%%%%%%%%%%%%%
\section{Comparison with BART}
Given the non-linear decision boundaries of the simulation scenario in Section 3.2, we further compare \TB with a standard non-linear Bayesian model. We use logistic Bayesian Additive Regression Trees (BART) as implemented in the BART package version 2.9 (lbart() function) \citep{bartpackage}.  

\begin{figure}[H]
	\begin{centering}
	\includegraphics[width = 0.8\textwidth]{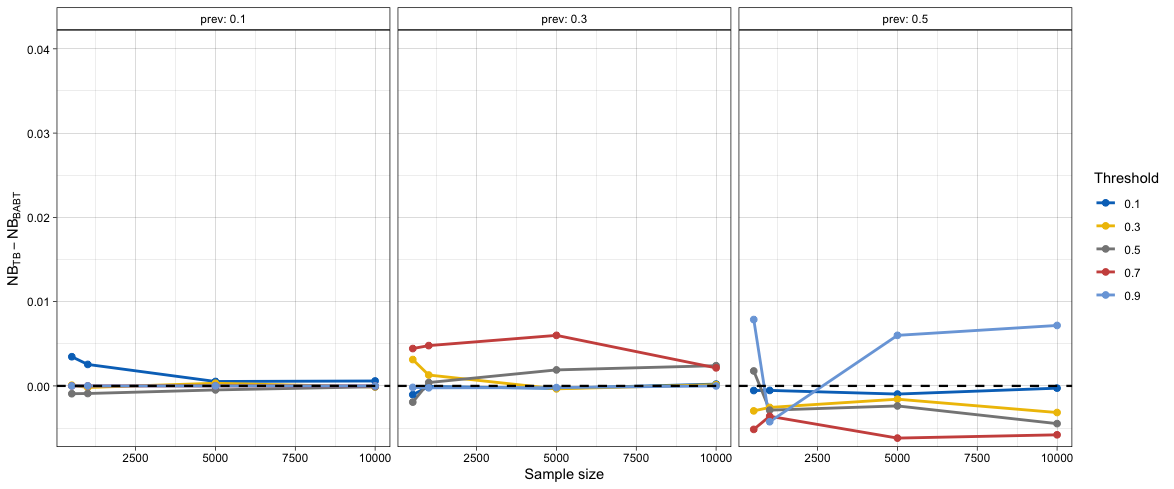}
    \caption{Difference in Net Benefit for samples sizes of 500, 1000, 5000, 10000 averaged over 20 repetitions. A positive difference means \TB outperforms BART. Each grid corresponds to a different prevalence setting.}
    \label{NB_TB_BART}
    \end{centering}
\end{figure}

Figure \ref{NB_TB_BART} shows the difference in NB between \TB and BART. Under the 0.5 prevalence scenario BART performs better than TB except at $t =0.9$. On the other hand, TB performs better or no worse than BART under prevalence scenarios 0.1 and 0.3. This is noteworthy as these prevalence scenarios are common in medical applications. Together with the results from Figure 4 (\manu), we conclude that for this simulation scenario, \TB, albeit implemented as a linear model, mitigates some of the advantages of a non-linear one, such as BART. Note that an additional comparison of interest would be BART with a tailored BART implementation. We leave this for future work.

%%%%%%%%%%%%%%%%%%%%%%%%%%%%%%%%%%%%%%%%%%%%%%%%%%%%%%%%%%%%%%%%%%%%%%%%%%%%%%%%%%
\section{Real data application 3: Breast cancer tumour classification}
\begin{figure}[H]
	\begin{centering}
	\includegraphics[width=0.8\textwidth]{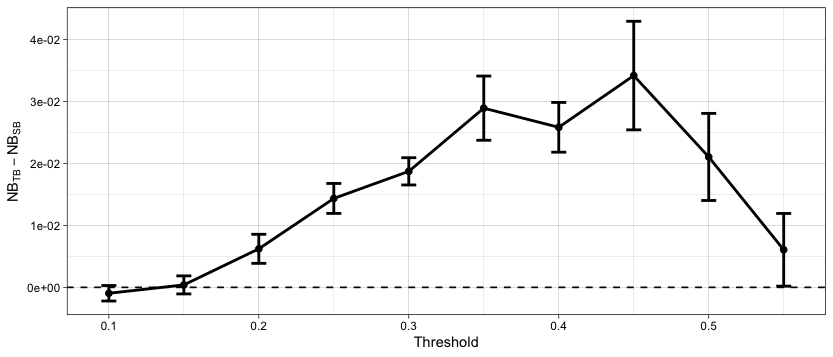}
    \caption{Difference in Net Benefit for various $t$ values evaluated on the test set. Error bars correspond to one standard error of the difference (see caption of Figure 7 \manu for details).}
    \label{fig1_real_data_3}
    \end{centering}
\end{figure}

For our third case study we use the Wisconsin breast cancer tumour dataset from the UCI repository \citep{Dua:2019}. The dataset consists of $n = 699$ points, with covariates $\mathbf{x} \in \mathbb{R}^{9}$, which describe characteristics of the cell nuclei present in digitized images of a breast mass, and labels $y \in \{0, 1 \}$. The class labels 0 and 1 correspond to ‘benign’ and ‘malignant’ cancers, respectively. To validate the results of the simulation in Section 3.3 in the \manu we artificially contaminate the dataset. More precisely, we use $70\%$ of the data for training, which is corrupted by flipping the labels of 49 class 1 datapoints to 0 ($10\%$ contamination). In clinical practice, such data contamination may arise due to the manual nature of breast cancer detection and classification. Breast cancer detection is commonly performed through medical imaging modalities by one or more experts (usually pathologists) \citep{murtaza2019deep}. The procedure is time-consuming and dependent on the professional experience and domain knowledge of the pathologists, thus making it prone to errors. This is highlighted by the significant inter- and intra-variability between pathologists  \citep{warfield2008validation, li2009variability, hong2012heterogeneity}. We use 20\% of the training data as design and the rest as development set. We assume that missing a malignant cancer is more severe than misdiagnosing a benign as malignant, and so we focus on target thresholds $t<0.5$, which correspond to a larger weight placed on false negatives vs false positives. Figure \ref{fig1_real_data_3} presents the difference in NB for various $t$ values over 5 splits of the training data into design and development. We see tailoring outperforms standard regression for most target thresholds.

\begin{figure}[H]
	\begin{centering}
	\includegraphics[width=1\textwidth]{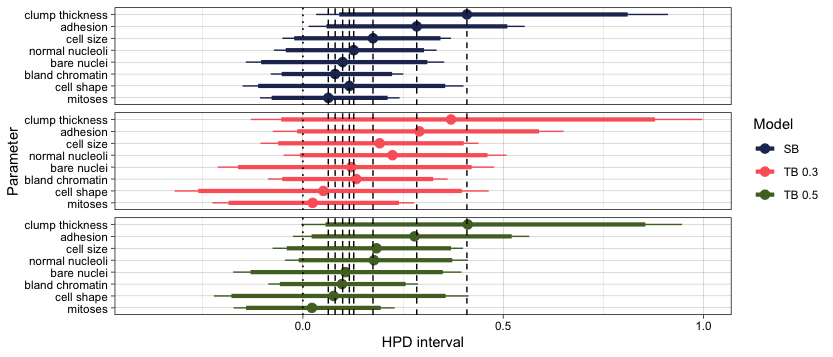}
    \caption{Highest posterior density (HPD) regions for the parameters. Dots represent medians, and thick and thin lines represent 90 and the 95\% of the HPD regions, respectively. The dashed vertical lines pass through the posterior median values of the \SB parameters.}
    \label{fig2_real_data_2}
    \end{centering}
\end{figure}

We further investigate the effect of tailoring on individual parameter values. Figure \ref{fig2_real_data_2} shows the highest posterior density (HPD) regions under \SB and \TB for $t= 0.3$ and $0.5$. As in the other case studies, under tailoring the regions are generally wider and are centred on different values. For instance, under $t= 0.3$ all posteriors are shifted towards more positives values. The only two exceptions are the coefficients of clump thickness, cell shape and mitosis which are pulled towards zero. Similar conclusions, but less pronounced are seen under $t = 0.5$.
This again indicates that the relative importance of different features changes when using our tailored modelling approach. 

%%%%%%%%%%%%%%%%%%%%%%%%%%%%%%%%%%%%%%%%%%%%%%%%%%%%%%%%%%%%%%%%%%%%%%%%%%%%%%%%%%
\section{Discussion Topics and Implementation Considerations}

The method presented in this paper relies on the construction of the datapoint-specific weights (see (7) in the \manu). Here we discuss each element in turn. 

\subsection{On choosing $\lambda$}
\begin{figure}[H]
	\begin{centering}
	\includegraphics[width = 0.8\linewidth]{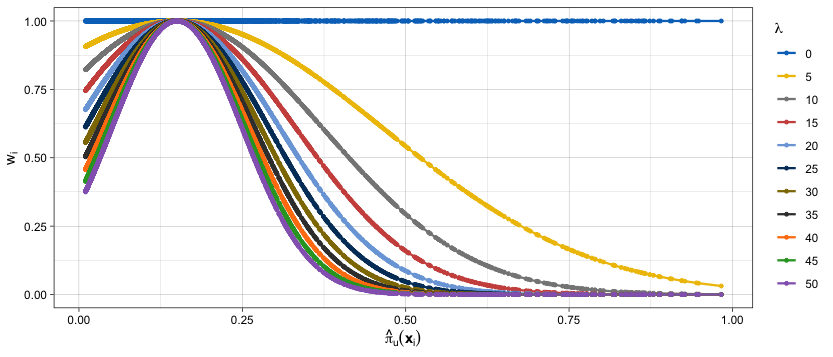}
    \caption{Distribution of weights, $w_i$, against $\hat{\pi}_u(\mathbf{x}_i)$ for breast cancer prognostication case study (Section 4.1 \manu) for $t = 0.15$}
    \label{weights_distribution}
    \end{centering}
\end{figure}

We have opted to use cross-validation to choose $\lambda$. An open question is how to choose the range of $\lambda$ values to consider. Our proposal is to consider values of the form $\lambda \in \{0,\dots, m\}$. When $\lambda = 0$, the model reduces to standard logistic regression, a sensible choice for the lower limit. To choose the upper limit, $m$, note that as $\lambda$ increases the rate with which the datapoints are downweighted increases exponentially (Figure \ref{weights_distribution}). This in turn decreases the effective number of datapoints that are used when training the model. We call this the effective sample size for tailoring, $ESS_T$. Formally, we define $ESS_T$ as 
\begin{equation*}
    ESS_T = \sum_{i=1}^n w_i.
\end{equation*}

\begin{figure}[H]
	\begin{centering}
	\includegraphics[width = 0.8\linewidth]{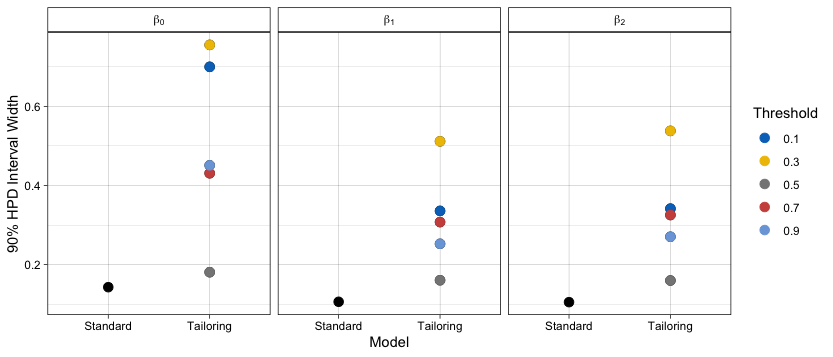}
    \caption{90\% HPD Interval width for each parameter as a function of the model.}
    \label{HPD_width_interval}
    \end{centering}
\end{figure}

Under standard modelling, $ESS_T = n$, since $w_i = 1$, $\forall i$. Under tailoring $ESS_T \le n$, which is why tailoring results in wider posteriors. This is demonstrated in Figure 1 of the main manuscript. In addition, Figure \ref{HPD_width_interval} shows the precision (as measured by the width) of the HPD credible intervals produced by each model under the simulation setting in Section 3.2 of the main manuscript. The figure suggests that the width of the credible intervals increases under tailoring compared to standard modelling. This is expected due to the downweighting of the likelihood contributions.

\begin{figure}[H]
	\begin{centering}
	\includegraphics[width = 0.8\linewidth]{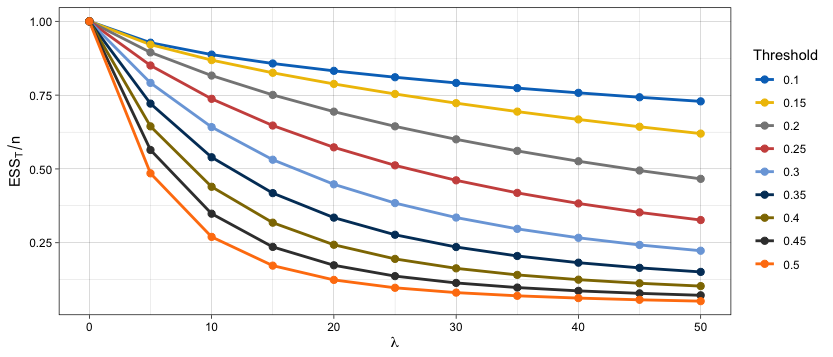}
    \caption{$\frac{ESS_T}{n}$ for various $\lambda$ values per target threshold.}
    \label{ess}
    \end{centering}
\end{figure}

As a result, we can use the $ESS_T$ as a guide to choose $m$. Figure \ref{ess} shows $\frac{ESS_T}{n}$ for various $\lambda$ values and target thresholds for the breast cancer prognostication case study (Section 4.1 \manu). Based on a target threshold we can choose $m$ so the $ESS_T$ does not drop below a pre-specified threshold. Importantly, this plot can be produced before fitting the model since we only need estimates of $\pi_u(\mathbf{x}_i)$. 

\begin{figure}[H]
	\begin{centering}
	\includegraphics[width = 0.8\linewidth]{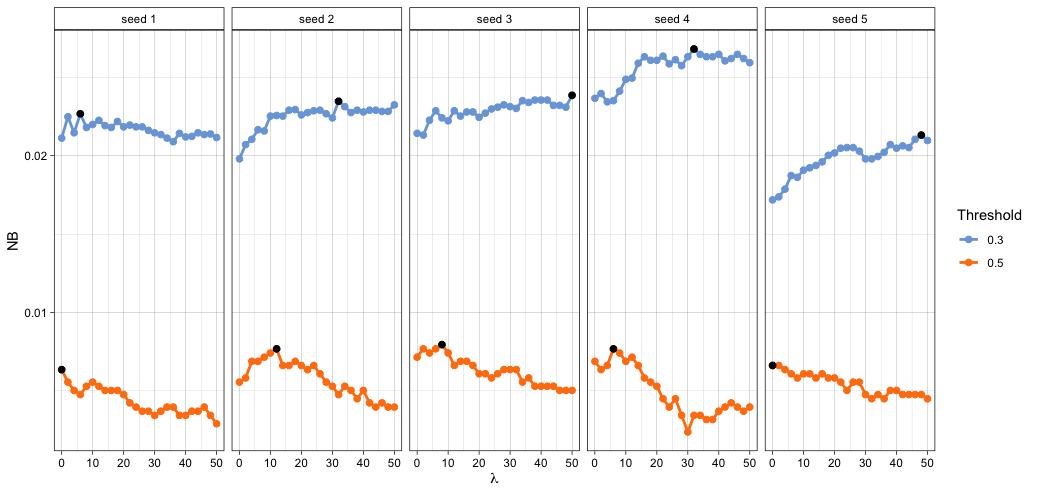}
    \caption{Average 5-fold CV estimate of Net Benefit (NB) for the breast cancer prognostication dataset. Black points correspond to the chosen lambda values, $\lambda^*$ (defined in Section \ref{choosing_lambda}).}
    \label{NB_lambda}
    \end{centering}
\end{figure}

In a similar fashion, we can have an indication whether \TB will outperform \SB before fitting the model. This can be achieved by plotting NB as $\lambda$ increases (Figure \ref{NB_lambda}). If NB remains stable or decreases as $\lambda$ increases (Figure \ref{NB_lambda}, $t = 0.5$ orange points) then \TB will probably not offer any performance improvement compared to \SB (see Figure 7 main text, $t= 0.5$). This is because, as discussed above, when $\lambda = 0$ \TB reduces to \SB (\ie all weights are equal to one).

%%%%%%%%%%%%%%%%%%%%%%%%%%%%%%%%%%%%%%%%%%%%%%%%%%%%%%%%%%%%%%%%%%%%%%%%%%%%%%%%%%
\subsection{On calibration}
Accurate estimation of $\pi_u(\mathbf{x}_i)$ at the first step of our framework is important for the construction of the weights.
Ideally, we would like the estimated probabilities, $\hat{\pi}_u(\mathbf{x}_i)$ to be well calibrated. Calibration refers to the degree of agreement between observed and estimated probabilities. Probabilities are well calibrated if, for every 100 patients given a risk of x\%, close to x have the event. 

\begin{figure}[H]
	\begin{centering}
		\subfloat[a][]{\includegraphics[width=0.5\linewidth]{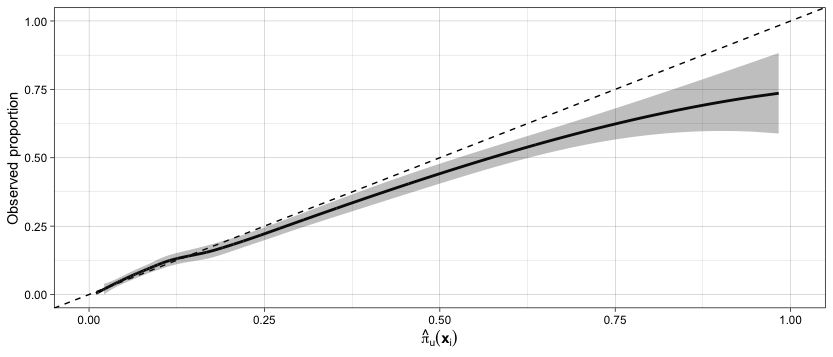}\label{fig_calibration:a}} 
		\subfloat[b][]{\includegraphics[width=0.5\linewidth]{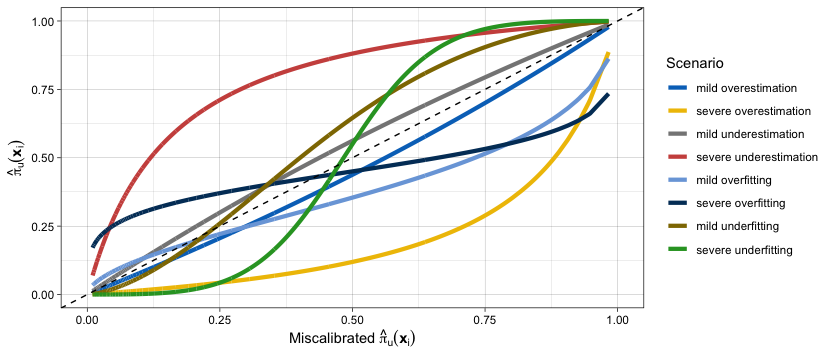}\label{fig_calibration:b}}
		\caption{(a) Calibration plot of $\hat{\pi}_u(\mathbf{x}_i)$ on the train data using loess smoother. The 45-degree line represents the perfect calibration. (b) Illustrations of different miscalibration scenarios. The y axis shows $\hat{\pi}_u(\mathbf{x}_i)$ and the x axis the miscalibrated $\hat{\pi}(\mathbf{x}_i)$ used in model fitting.} 
		\label{calibration}
    \end{centering}
\end{figure}

We use the breast cancer prognostication case study (Section 4.1 \manu) to investigate the effect of miscalibration on the model performance. First, we assess the calibration of $\hat{\pi}_u(\mathbf{x}_i)$. Figure \ref{fig_calibration:a} presents a graphical evaluation of calibration. It is based on loess-based smoothing method \citep{austin2014graphical} where the estimated (\ie $\hat{\pi}_u(\mathbf{x}_i)$) and observed probabilities (from the development data) are plotted against each other; good models are close to the 45-degree line. We see the probabilities are well calibrated for the lower risks, and tend to be underestimated for the higher risks. 
To explore sensitivity of the tailored model to the accuracy of the step 1 probabilities, we deliberately perturbed $\hat{\pi}_u(\mathbf{x}_i)$ generating four miscalibration types: (1) overestimation; (2)  underestimation, when probabilities are systematically overestimated or underestimated, respectively;  (3) overfitting, when small probabilities are underestimated whereas large ones are overestimated; (4) underfitting, when small probabilities are overestimated whereas large ones are underestimated. We further allowed for two degrees of miscalibration (mild and severe) for each type giving us a total of eight scenarios (Figure \ref{fig_calibration:b}). 

\begin{figure}[H]
	\begin{centering}
	\includegraphics[width = 0.9\textwidth]{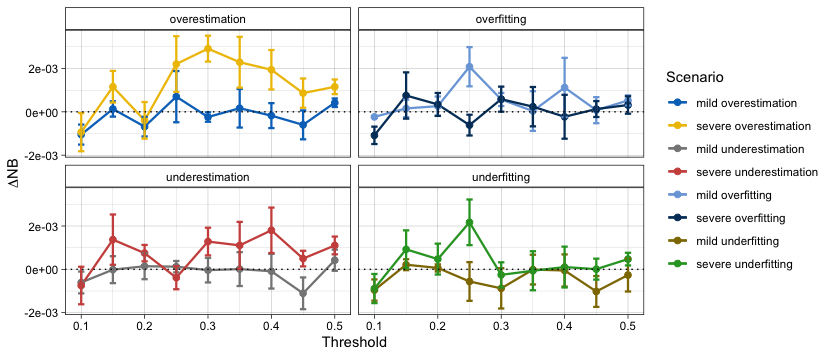}
    \caption{Difference in NB, $\Delta$NB (on the test set) between original, calibrated \TB and miscalibrated \TB under different scenarios. A positive difference means the calibrated \TB outperforms the miscalibrated one.}
    \label{miscalibration}
    \end{centering}
\end{figure}

Figure \ref{miscalibration} shows the difference in NB between the original tailored (calibrated) model and the tailored miscalibrated ones for the different scenarios. Comparing across miscalibration types we see that the decline in performance depends on the type of miscalibration, with overfitting and underfittng more robust than over- and underestimation. Comparing within each type we note a drop in performance from mild to severe degrees, especially for over- and underestimation. 

These results show that the model performance (in terms of NB) depends on the type of miscalibration and is robust to mild miscalibration. In practice, the calibration of the estimated probabilities can be readily evaluated graphically as done here or using statistical tests \citep{austin2014graphical}. If the results show poor calibration we recommend re-calibrating $\hat{\pi}_u(\mathbf{x}_i)$ before calculating the datapoint-specific weights  \citep{steyerberg2004validation, janssen2008updating}. 

%%%%%%%%%%%%%%%%%%%%%%%%%%%%%%%%%%%%%%%%%%%%%%%%%%%%%%%%%%%%%%%%%%%%%%%%%%%%%%%%%%
\subsection{On the weighting function}
In Section 2.3 of the \manu we defined the weights using the squared distance function, $h$. Here we investigate the sensitivity of the framework to the choice of the distance function. We choose the family of $\epsilon$-insensitive functions \citep{vapnik1998statistical}, which is defined as
$$ h(\pi_u(\mathbf{x}), t) = |\pi_u(\mathbf{x}) - t|_{\epsilon}$$

where we denote 
\begin{equation}
%\[
 |\pi_u(\mathbf{x}) - t|_{\epsilon} = 
  \begin{cases} 
   0 		& \text{if } |\pi_u(\mathbf{x}) - t| \leq \epsilon \\
   |\pi_u(\mathbf{x}) - t|	- \epsilon	& \text{otherwise }   
  \end{cases}
%\]
\label{epsilon_insensitive}
\end{equation}

\begin{figure}[H]
	\begin{centering}
	\includegraphics[width = 0.9\textwidth]{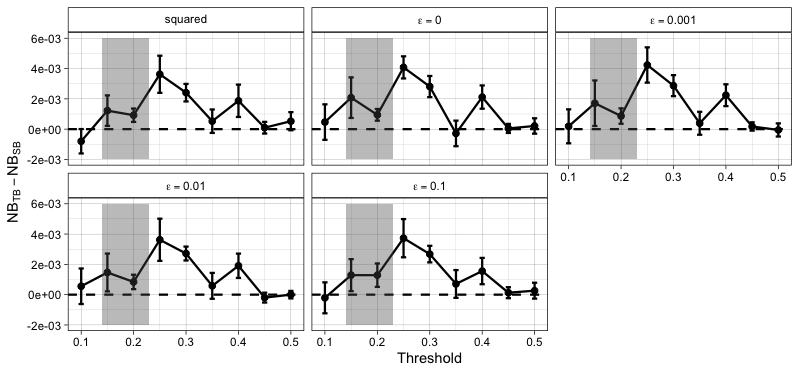}
    \caption{Difference in NB (breast cancer prognostication case study) between \TB and \SB 
    under the squared distance and $\epsilon$-insensitive functions for 
    various $\epsilon$ values. Note the first panel corresponds 
    to Figure 7 in the \manu.}
    \label{weighting_function}
    \end{centering}
\end{figure}

The $\epsilon$-insensitivity arises from the fact that the function value is equal to 0 if the discrepancy between the predicted probability $\pi_u(\mathbf{x})$ and the target threshold $t$ is less than $\epsilon$. In other words, we do not care about the distance as long as it is less than $\epsilon$, but will not accept any deviation larger than this. As a result, observations with predicted probability within $\epsilon$ of the target threshold will not be downweighted. For $\epsilon = 0$ we recover the absolute distance, which is the objective function in median regression \citep{bassett1978asymptotic}. Both the squared distance and the family of $\epsilon$-insensitive functions are symmetric, \ie they downweight equally observations based only on their distance from the target threshold, not taking into account the direction. This is a reasonable requirement for our weighting function. 
Figure \ref{weighting_function} presents the results for various $\epsilon$ values for the breast cancer case study (Section 4.1 \manu). The conclusions are qualitatively unchanged when compared within different $\epsilon$ values and between $\epsilon$-insensitive functions and the squared distance (first panel in Figure \ref{weighting_function}). Hence, we conclude that for this dataset the results are also robust to the choice of the weighting function. 

%%%%%%%%%%%%%%%%%%%%%%%%%%%%%%%%%%%%%%%%%%%%%%%%%%%%%%%%%%%%%%%%%%%%%%%%%%%%%%%%%%
%\bibliographystyle{biorefs}
%\bibliography{suppl/refs2}

%\end{document}

\end{document}